\documentstyle{mn}

\begin{document}
                                                                                
\title[Spectroscopy of two overlapping samples of radio sources]
{Optical spectroscopy of two overlapping, flux-density--limited samples of 
radio sources in the North Ecliptic Cap, selected at 38 MHz and 151 MHz}

\author[Lacy et al.]{Mark Lacy$^{1}$ 
\thanks{email: m.lacy1@physics.oxford.ac.uk}, Steve Rawlings$^{1}$, Gary
J. Hill$^2$, Andrew J.\ Bunker$^{3}$, Susan E.\ Ridgway$^{1,4}$ \\
{\LARGE \& Daniel Stern$^{3}$} \\
$^{1}$Astrophysics, Department of Physics, Keble Road, Oxford, OX1 3RH.\\
$^{2}$McDonald Observatory, University of Texas at Austin, TX78712-1083\\
$^{3}$Dept of Astronomy, University of California at Berkeley, 601 Campbell 
Hall, Berkeley CA 94720 \\
$^{4}$Bloomberg Center for Physics \& Astronomy, Johns Hopkins University, 
3400 Charles Street, Baltimore MD 21218\\}

\maketitle
\begin{abstract}
We present the results of optical spectroscopy of two flux-density--limited 
samples of radio sources selected at frequencies of 38 and 151 MHz in the 
same region around the North Ecliptic Cap, the 8C-NEC and 7C-{\sc iii} samples 
respectively. Both samples are selected at flux density levels $\approx 20$ 
times fainter than samples based on the 3C catalogue. They are amongst the 
first low-frequency selected samples with no spectral or angular size 
selection for which almost complete redshift information has been obtained, 
and contain many of the lowest-luminosity $z>2$ radio galaxies so far 
discovered. They will therefore provide a valuable resource for understanding 
the cosmic evolution of radio sources and their hosts and environments. The 
151-MHz 7C-{\sc iii} sample is selected to have $S_{151} \geq 0.5$Jy and is 
the more spectroscopically complete; out of 54 radio sources fairly reliable
redshifts have been obtained for 44 objects. The 8C sample has a flux limit of 
$S_{38} \geq 1.3$Jy 
and contains 58 sources of which 46 have fairly reliable redshifts.
We discuss possible biases in the observed redshift distribution, 
and some interesting individual objects, including a number of cases of 
probable gravitational lensing. Using the 8C-NEC and 7C-{\sc iii} 
samples in conjunction, we form the first sample selected on 
low-frequency flux in the rest-frame of the source, 
rather than the usual selection on flux density in the observed frame.
This allows us to remove the bias associated with an increasing rest-frame 
selection frequency with redshift. We investigate the difference this 
selection makes to correlations of radio source properties with redshift and 
luminosity by comparing the results from traditional flux-density selection
with our new method. We show in particular that flux-density--based selection
leads to an overestimate of the steepness of the correlation of radio 
source size with redshift.  
\end{abstract}


\section{Introduction}

Spectroscopic surveys of radio sources at flux density levels much lower than 
those of the 3C catalogue and its various revisions (e.g.\ Laing, Riley
\& Longair 1983, hereafter LRL) are 
essential to provide coverage of the redshift-radio luminosity plane. Such 
coverage is required to establish the cosmic evolution of these sources both 
in number density and physical properties such as size and structure. 
Over the past two decades, several flux-limited samples selected at low radio 
frequency (and hence directly comparable with LRL in being selected primarily 
on the basis of unbeamed steep spectrum radio lobe emission) have been defined
and imaged in the radio and 
optical. Only recently, however, have redshifts have been obtained for a 
significant fraction of the radio sources. This paper describes spectroscopic
observations of the 38-MHz 8C North Ecliptic Cap (NEC) sample
of Lacy, Rawlings \& Warner (1992; hereafter Paper I) and Lacy et al.\ 1993,
1999a (hereafter Papers II and III respectively), and the closely-related 
7C-{\sc iii} sample. This latter sample was selected at a frequency of 151 MHz
using the 7C survey data of Lacy et al.\ (1995) with a flux limit chosen to 
maximise overlap with the 8C sample. This sample is defined in 
Section 2, and will be 
described in more detail in a future paper (Lacy et al.\ 1999b). 
A further paper will discuss spectroscopy of the 6C sample of Eales (1985), 
and details of the 7C-{\sc i} and 7C-{\sc ii} samples are presented in Willott
(1998, and papers in preparation). The properties of these samples
are summarised in Table 1, and the coverage of the radio-luminosity -- 
redshift plane is illustrated in, e.g.\  fig.\ 1 of Blundell, 
Rawlings  \& Willott (1999). 

Samples such as these are very valuable for studying the cosmic 
evolution of the radio source population, in particular the question of
the evolution of source size, $D$, with redshift. 
%Most observational work on the evolution of size with redshift has been 
%done using the Spearman rank coefficients to perform a non-parametric 
As redshift and radio luminosity are always strongly correlated within any 
flux-limited sample,
partial correlation coefficients need to be calculated to investigate 
whether the true correlation is with redshift or luminosity. Studies based on
low-frequency selected samples typically show a weak correlation with redshift
only, parameterised as $D\propto (1+z)^{-\eta}$ where $\eta \approx 
1.7-1.9$ (e.g.\ the 6C sample, Eales 1985; Neeser et al.\ 1995, and the 
7C-{\sc i} and 7C-{\sc ii} samples; Blundell, Rawlings \& Willott 1999), 
although samples selected at higher frequencies seem to have 
stronger dependences, with $\eta \approx 3$ 
(Oort, Katgert \& Windhorst 1987; Kapahi 1989) combined with a radio 
luminosity dependence $D\propto L^{\epsilon}$ with $\epsilon \approx 0.3$. 

In Paper II we performed a partial rank correlation on the 8C-NEC sample 
using photometric redshifts. This showed that the fundamental 
correlation was between size and redshift, but having only redshifts based
on $R$-band magnitudes we were forced to only consider $z<1.1$ objects and 
to accept the uncertainties associated with photometric redshifts. We also 
investigated the four-way correlation coefficients of redshift, luminosity, 
size and spectral index, finding that spectral index at 2GHz in the rest frame
was most strongly correlated with redshift, and any correlation with 
luminosity was weak. Since then, independent work on the 6C, 7C-{\sc i} and 
7C-{\sc ii} surveys has been carried out by Blundell et al.\ 
(1999) with results that are broadly consistent with 
the Neeser et al.\ result and the results of Paper II. They find, however,
a significant correlation of spectral index with luminosity at a rest-frame 
frequency of 1 GHz, although a correlation with redshift is seen at higher 
frequencies. In this paper we use the redshifts for the 7C-{\sc iii}
and 8C-NEC samples to study the size-redshift relation further.

We assume $H_0=50 \, {\rm kms^{-1}Mpc^{-1}}$ and an Einstein --
de Sitter ($\Omega_{\rm M} = 1$, $\Omega_{\Lambda}=0$) cosmology unless 
otherwise stated. Positions are all epoch B1950.0. Spectral indices, both 
radio and optical are defined in the sense that 
$S_{\nu} \propto \nu^{-\alpha}$.

\begin{table}
\caption{Complete, flux-density limited radio source samples at low radio
frequency selected from the Cambridge surveys for which optical spectra have 
been obtained}
\begin{tabular}{ll}
Name  & Selection criteria \\
              &                    \\
6C   & $2<S_{151}<4$ Jy,\\ 
     & $8^{\rm hr}20^{\rm m}<{\rm R.A.}<13^{\rm hr} 01^{\rm m}$, 
$34^{\circ}< {\rm Dec.} < 40^{\circ}$\\
7C-{\sc i} & $S_{151}\geq 0.5$ Jy  in the 5C6 field \\
7C-{\sc ii} & $S_{151}\geq 0.5$ Jy  in the 5C7 field \\
7C-{\sc iii}&54 sources with $S_{151}\geq 0.5$ Jy within $3^{\circ}$ of the NEC\\ 
       & [for maximum overlap with the 8C NEC sample]\\
8C-NEC  &58 sources with $S_{38}\geq 1.3$ Jy within $3^{\circ}$ of the NEC\\
%6CS  & $1.0<S_{151}<2$ Jy;  $??<{\rm R.A.}<?$; $\alpha^{151}_{4850}>1.2$; 
%$\theta < 10$ arcsec \\
\end{tabular}
\end{table}

\section{Sample definitions and the spectroscopic sub-samples}

The 8C-NEC sample has been formally defined in Paper I, and some refinements 
described in Paper III. It consists of all radio sources with 38-MHz flux
densities 
$S_{38} \geq 1.3$ Jy within $3^{\circ}$ of 18$^{\rm h}$00$^{\rm m}$ 
+66$^{\circ}$ in the survey of Rees (1990). 
The 7C-{\sc iii} sample consists of all 7C 
objects from the mini-survey of Lacy et al.\ (1995) with 151-MHz flux 
densities 
$S_{151}\geq 0.5$ Jy, again within $3^{\circ}$ of $18^{\rm h} 00^{\rm m}$ 
$+66^{\circ}$. The flux density limits are deliberately chosen to maximise the 
overlap of the two samples, thus most sources are common to both samples.
The 7C-{\sc iii} sample will be more fully described in 
Lacy et al.\ 1999b, but the positions and radio properties of the objects which
are present only in the 7C-{\sc iii} sample (and which therefore have 
not appeared in Papers I-III) are presented in Table 2, for 
completeness. A number of objects have 
been excluded from the samples on the basis of confusion by bright nearby
sources. 7C 1732+6715 (8C 1732+672) has a very bright radio source nearby
and has been temporarily removed from the sample pending an improved 
radio image. 7C 1821+6442 (8C 1821+646) has a star on top of the ID position
(R.G.\ McMahon, personal communication).
7C 1827+6517 (8C 1827+652) has been excluded as there is a very bright star
nearby. This has reduced the total number of objects in the 7C-{\sc iii} 
sample from 
57 to 54 and for the 8C-NEC sample from 61 to 58. As these objects were
removed for reasons unconnected with their intrinsic radio or optical 
properties their omission should not affect the sample statistics. Apart from 
these, all objects in the 7C-{\sc iii} sample have spectroscopic data.
In the 8C-NEC sample, there are three objects for which spectra are yet to be
obtained.

\begin{table*}
\caption{Objects in the 7C-{\sc iii} sample which are not in the 8C-NEC sample}
\begin{tabular}{lcccrrrrl}
Name       & $S_{151}$ & R.A.\ (1950) & Dec.\ (1950) &$\alpha^{38}_{151}$& $R$ & Radio &Radio& Notes \\
           & /Jy       &              &              &  & & size/$\; ^{''}$&PA
\\
7C1731+6641&0.52&17 31 43.57&+66 40 59.3&0.40& 21.4& 1.0 & 164&\\
7C1740+6640&0.54&17 40 42.80&+66 40 07.2&0.37& 24.8&$<$0.5& - &see Section 4.3\\
7C1745+6624&0.51&17 45 57.20&+66 24 18.5&0.49&  -  &$<$0.5& - &see Section 4.3\\
7C1748+6657&1.15&17 48 17.57&+66 57 15.8&$<$-0.26& 22.4&0.3   & 90&\\
7C1807+6719&0.71&18 07 19.56&+67 19 11.6&0.25&$>$23& 1.9  & 47&\\ 
7C1812+6814&0.59&18 12 15.65&+68 14 06.5&0.51&$>$23&22.0  &123&\\
7C1820+6657&0.83&18 20 40.21&+66 57 12.1&$<$-0.03&$>$23&$<$0.5& - &\\
\end{tabular}

\vspace*{0.1in}

\noindent
Note: positions are those of the optical/IR identification except for 
7C1812+6814 where the identification is only seen in the spectrum and 
7C1820+6657 which is unidentified optically apart from a very faint 
emission line in the spectrum. In these cases the radio 
position has been quoted (this is the mid-point of the hotspots in the case 
of 7C1812+6814). 
\end{table*}

\section{Spectroscopic Observations}

Most of the objects in both the 8C-NEC and 7C-{\sc iii} North Ecliptic Cap 
samples were observed with the ISIS 
spectrograph on the William Herschel Telescope (WHT). Most observations were 
made on the nights of 1995 July
28 to 31, though observations of a few objects were made by on the nights
of 1993 August 20-21 and 1993 June 18-19. 
Observations of brighter objects in the sample
were made with the IGI on the McDonald Observatory 107$^{''}$ telescope 
on the nights of 1993 July 27-28. Two objects were observed 
using the Kast spectrograph on the Shane 3-m telescope at the Lick 
Observatory. Details of the observations are given in Tables 3 and 5.

In most cases, a long spectroscopic slit was centred on the optical 
identification and aligned with the radio axis, offsetting from a 
nearby star (positions of offset stars are available from ML). In a few 
cases there
was no identification on optical or infrared images, in which case the 
spectrum was taken blind, with the slit aligned along the radio axis.

The ISIS observations were made using both arms and a dichroic at either 
540nm (in 1993) or 570nm (in 1995). In 1993 June and August the red and 
blue arm detectors were an EEV5 CCD and a TEK CCD respectively, in 1995 July 
both detectors were TEK CCDs. The TEK CCD suffers from fringing at the
red end of the red arm, but was preferred to the EEV5 in 1995 because of 
its higher quantum efficiency. In practice, it proved possible to remove the 
fringing effectively by taking tungsten lamp flats immediately following 
each observation, before moving the telescope. 

The IGI observations were made as described in Paper II. The
Kast observations were made with both red and blue arms, the beam being
split by a dichroic at 550 nm. Reticon CCDs were used as detectors in 
both arms and 452 and 300
lmm$^{-1}$ gratings were employed in the blue and red arms
respectively. 

Data reduction of the WHT, 107$^{''}$ and Lick data followed the standard 
procedure of bias subtraction, division by a flatfield, wavelength
calibration using arc lamp spectra (CuAr and CuNe for the WHT data, 
Cd for the 107$^{''}$ data and CdHeHg and NeAr for the Lick data) 
and flux calibration using spectrophotometric standard stars. 

Errors in the spectrophotometry are expected to be $\approx 15$ per cent; 
errors in the wavelengths of the bright spectral features are not expected
to be larger than 0.2 nm in the WHT and Lick data and 0.3 nm in the McDonald 
data.

\subsection{The 7C-{\sc iii} spectra}

In Fig.\ 1 we present the spectra of the objects in the 7C-{\sc iii} sample
with the main spectral features 
labelled. The WHT spectra in the figure 
have been smoothed with a 1.5 nm box-car filter, and the
McDonald and Lick spectra by a 1.2 nm one. If features appear
in only one arm of the WHT spectrum, the other has not been shown.
All spectra for objects in the spectroscopic sub-sample
are shown except for those already published in Paper II, or Lacy et al.\ 
(1999c; hereafter LRWLR), those for which spectra have been published 
by Kollgaard et al.\ (1995; hereafter K95) and those for which 
there are no spectral features visible. These are shown separately in 
Fig.\ 2. Derived redshifts 
and emission line properties (flux, equivalent width and velocity
width) are listed in Table 4. The emission 
line properties are derived from Gaussian fits to the data. 

\subsection{Spectra of objects in the 8C-NEC sample only}

We have also obtained data on some objects in the original 8C-NEC sample 
of Paper I which were just below the $S_{151}=0.5$Jy flux
density cutoff of the 7C-{\sc iii} sample. The observations are listed in Table
5 and the redshifts and other emission line properties in Table 6,
with the whole sample listed for completeness. The spectra are shown
in Fig.\ 6, smoothed as in Fig.\ 1 except for 8C 1758+676 which appears to 
be a faint quasar, and to which a smoothing of 4 nm has been applied to 
bring out the broad lines.

\begin{table*}
\caption{Log of new spectroscopy of the 7C-{\sc iii} radio sources}
\begin{tabular}{lcclrcrr}
7C Name & Observing date(s)&Telescope & wavelength range(s) & integration
& slit width & slit PA& airmass\\
     &               &          &                 &  time(s)/sec & /arcseconds
 & \\
7C 1731+6641&31/07/95&WHT&330 - 850 nm&500R, 650B&3 & 164&1.27\\
7C 1732+6715&28/07/95&WHT&330 - 850 nm&300       &2 & 200&1.29\\
%7C 1732+6535 &&&&&&\\
7C 1733+6719&29/07/95&WHT&330 - 850 nm&1800      &3 &15  &1.32\\
%7C 1736+6710& &107''&&&&\\
7C 1740+6640&29/07/95&WHT&330 - 850 nm&1800      &3&113&1.55\\
            &31/07/95&WHT&330 - 850 nm&1800&3&20&1.34\\
7C 1741+6704&20/08/93&WHT&330 - 850 nm&1800B, 1700R&3&56&1.44\\
            &        &   &            &            & &  &    \\
7C 1742+6346&28/07/95&WHT&330 - 850 nm&1800&3&63&1.22\\
7C 1743+6431&28/07/95&WHT&330 - 850 nm&1800&3&90&1.53\\ 
            &29/07/95&WHT&330 - 850 nm&1800&3&88&1.25\\
% above also done night 2?
7C 1743+6344&30/07/95&WHT&330 - 850 nm&1800&3&10&1.47\\
%7C 1743+6639& &107''&&&&\\
7C 1745+6415&31/07/95&WHT&330 - 850 nm&500R, 560B&3&177&1.24\\
7C 1745+6422&30/07/95&WHT&330 - 850 nm&1280R, 1330B&3&35&1.23\\
            &        &   &            &            & &  &    \\
7C 1745+6624&29/07/95&WHT&330 - 850 nm&1800&3&105&1.65\\
%7C 1747+6533& &&&&&\\
7C 1748+6703&30/07/95&WHT&330 - 850 nm&1800&3&90&1.29\\
7C 1748+6657&29/07/95&WHT&330 - 850 nm&1650R, 1700B&3&144&1.36\\
7C 1748+6731&29/07/95&WHT&330 - 850 nm&600&3&0&1.31\\
7C 1751+6809&30/07/95&WHT&330 - 850 nm&1800&3&31&1.33\\
            &        &   &            &            & &  &    \\
7C 1751+6455&27/07/93&107''&430 - 700 nm&3600&2&-&1.25\\
7C 1753+6311&19/08/93&WHT&330 - 850 nm&1800B, 1700R&3&74&1.60\\
            &31/07/95&WHT&330 - 850 nm&1800&3&70&1.24\\
%7C 1753+6543& &&&&&\\
7C 1754+6420&28/07/95&WHT&330 - 850 nm&1800&2.5&15&1.23\\
7C 1755+6314&28/07/93&107''&430 - 700 nm&3600&2&-&1.31\\
7C 1755+6830&31/07/95&WHT&330 - 850 nm&600&3&65&1.35\\
            &        &   &            &            & &  &    \\
7C 1756+6520&30/07/95&WHT&330 - 850 nm&1800&3&47&1.32\\
7C 1758+6535&29/07/95&WHT&330 - 850 nm&1800&3&52&1.26\\
7C 1758+6553&28/07/93&107''&430 - 700 nm&1800&2&-&1.52\\
7C 1758+6307&28/07/95&WHT&330 - 850 nm&900&3&91&1.84\\
7C 1758+6719&28/07/95&WHT&330 - 850 nm&1800&3&177&1.25\\
            &        &   &            &            & &  &    \\
%7C 1801+6902& &107''&&&&\\
7C 1802+6456&18/06/93&WHT&330 - 850 nm&1800&3&25&1.27\\
7C 1804+6625&28/07/95&WHT&330 - 850 nm&600&2&206&1.29\\
7C 1804+6313&31/07/95&WHT&330 - 850 nm&$2\times 600$&3&47,112&1.26,1.47\\
7C 1805+6332&20/08/93&WHT&330 - 850 nm&1800&3&0&1.55\\
7C 1807+6831&29/07/95&WHT&330 - 850 nm&600&3&60&1.32\\
            &        &   &            &            & &  &    \\
7C 1807+6719&29/07/95&WHT&330 - 850 nm&1800&3&52&1.26\\
7C 1807+6841&28/07/95&WHT&330 - 850 nm&1520R, 1600B&3&177&1.30\\
7C 1811+6321&29/07/95&WHT&330 - 850 nm&600&3&22&1.64\\
7C 1812+6814&29/07/95&WHT&330 - 850 nm&1800&3&123&1.38\\
7C 1813+6846&29/07/95&WHT&330 - 850 nm&300&2&0&1.37\\
            &        &   &            &            & &  &    \\
7C 1813+6439&05/07/97&Shane&330 - 1060 nm&3600&2&-&1.3\\
7C 1814+6702&20/08/93&WHT&330 - 900 nm&1800B, 1700R&3&156&1.36\\
            &31/07/95&WHT&330 - 850 nm&1800&2&156&1.36\\
7C 1814+6529&30/07/95&WHT&330 - 850 nm&1800&3&48&1.25\\
7C 1815+6815&17/06/93&WHT&330 - 850 nm&1700&3&100&1.31\\
%7C 1815+6805& &107''&&&&\\
7C 1816+6710&28/07/95&WHT&330 - 850 nm&1800&3&90&1.60\\
            &        &   &            &            & &  &    \\
7C 1816+6605&31/07/95&WHT&330 - 850 nm&1500B, 1450R&2&40&1.27\\
7C 1819+6550&30/07/95&WHT&330 - 850 nm&1650&3&70&1.51\\
7C 1820+6657&30/07/95&WHT&330 - 850 nm&1800&3&178&1.37\\
7C 1822+6601&28/07/93&107''&430 - 700 nm&3600&2&-&1.61\\
7C 1825+6602&31/07/95&WHT&330 - 850 nm&1800&2-3&78&1.26\\
            &        &   &            &            & &  &    \\
7C 1826+6510&28/07/95&WHT&330 - 850 nm&1800&3&123&1.36\\
%7C 1826+6704& &107''&&&&\\
7C 1827+6709&28/07/95&WHT&330 - 850 nm&1570B, 1500R&2.5&137&1.34\\
\end{tabular}
\end{table*}

\section{Results}

\begin{figure*}
\begin{picture}(400,500)
\put(-100,-190){\includegraphics{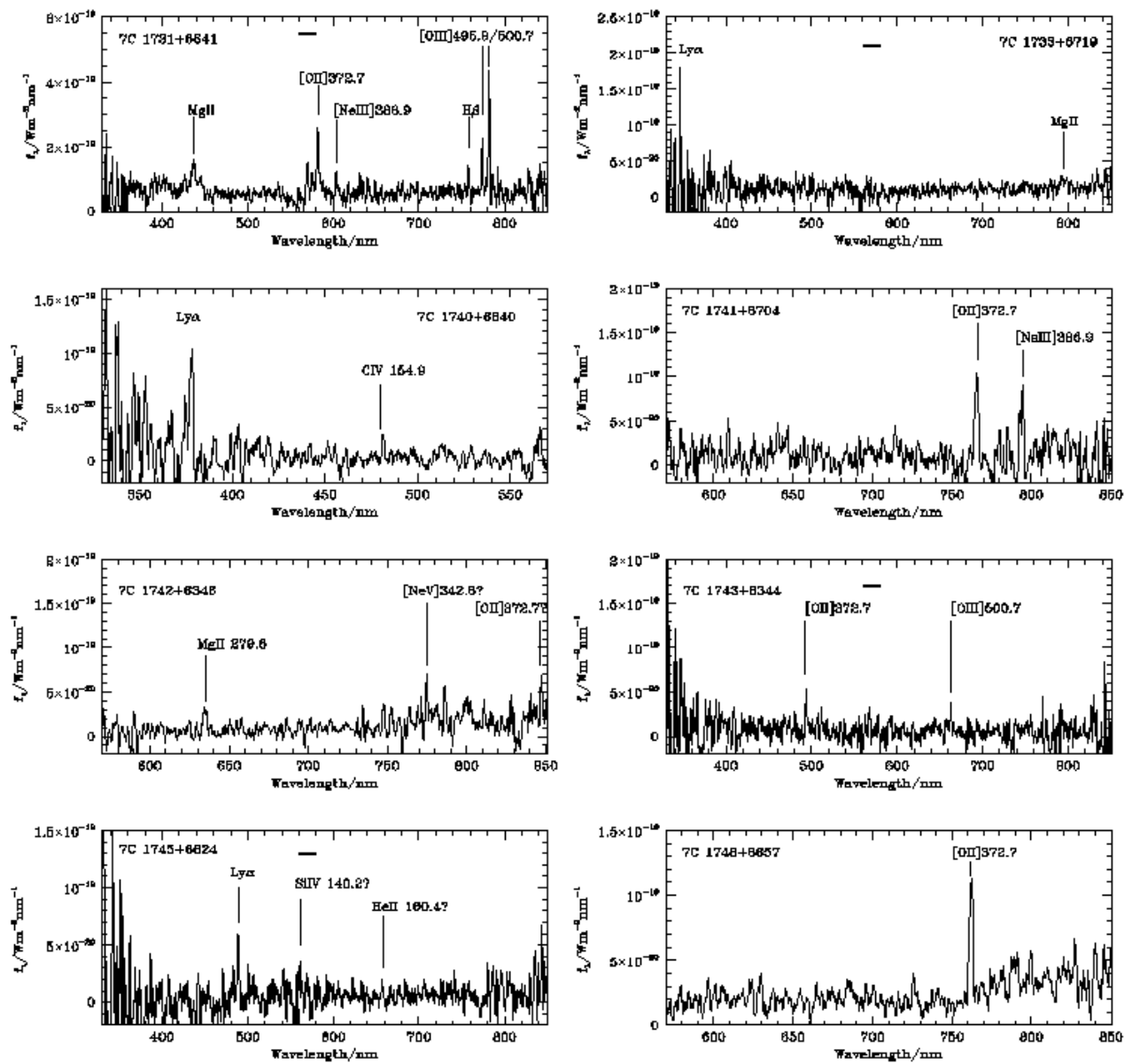}}
\end{picture}
\caption{Spectra of objects in the 7C-{\sc iii} sample. The thick black bars
on some of the plots show the wavelength region where the dichroic 
transmission was $<<1$ on the crossover from the red to blue arms. This region
is noisy and sensitive to inaccuracies in the calibration.}
\end{figure*}

\setcounter{figure}{0}

\begin{figure*}
\begin{picture}(400,500)
\put(-100,-190){\includegraphics{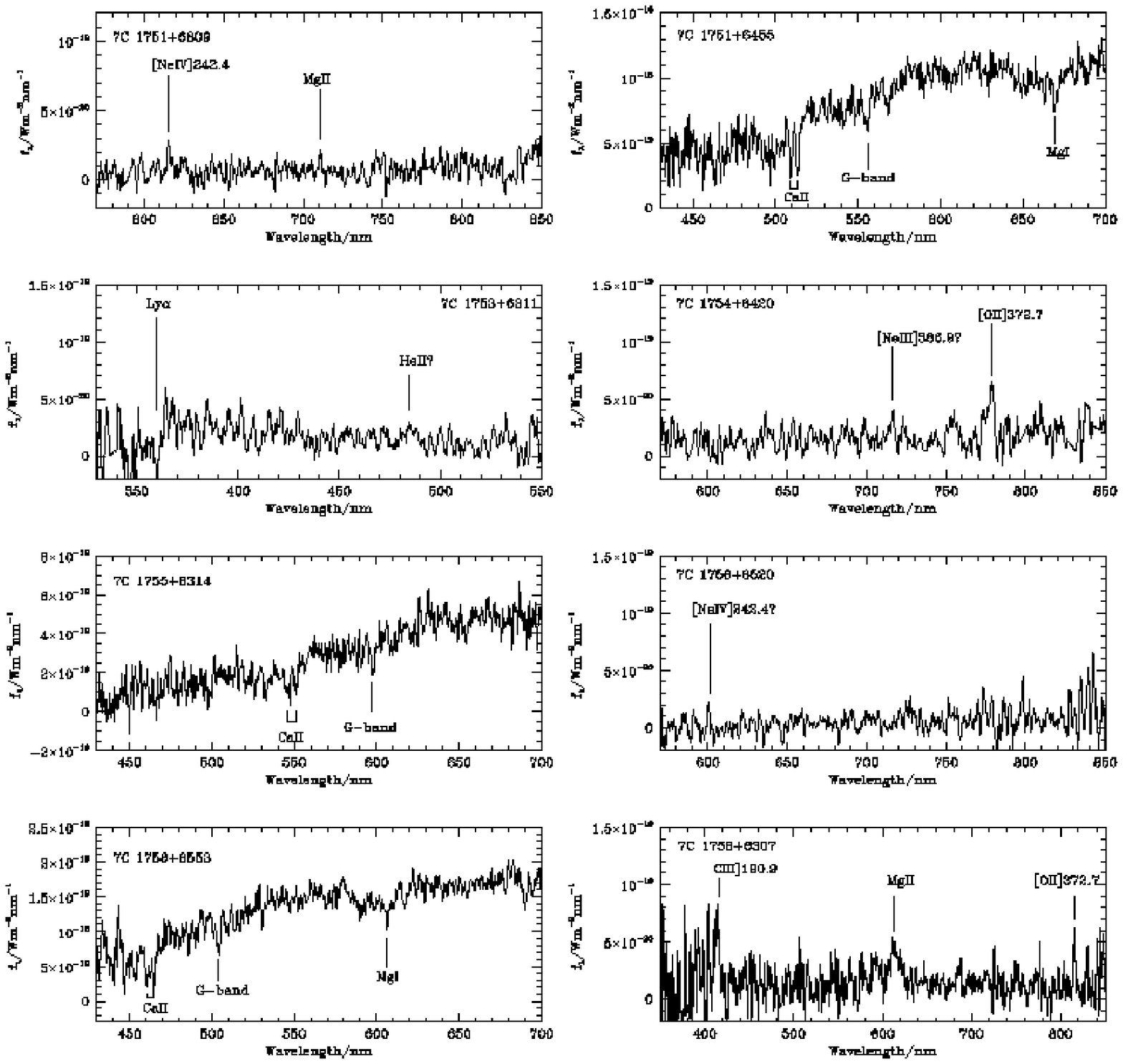}}
\end{picture}
\caption{continued}
\end{figure*}

\setcounter{figure}{0}

\begin{figure*}
\begin{picture}(400,500)
\put(-100,-190){\includegraphics{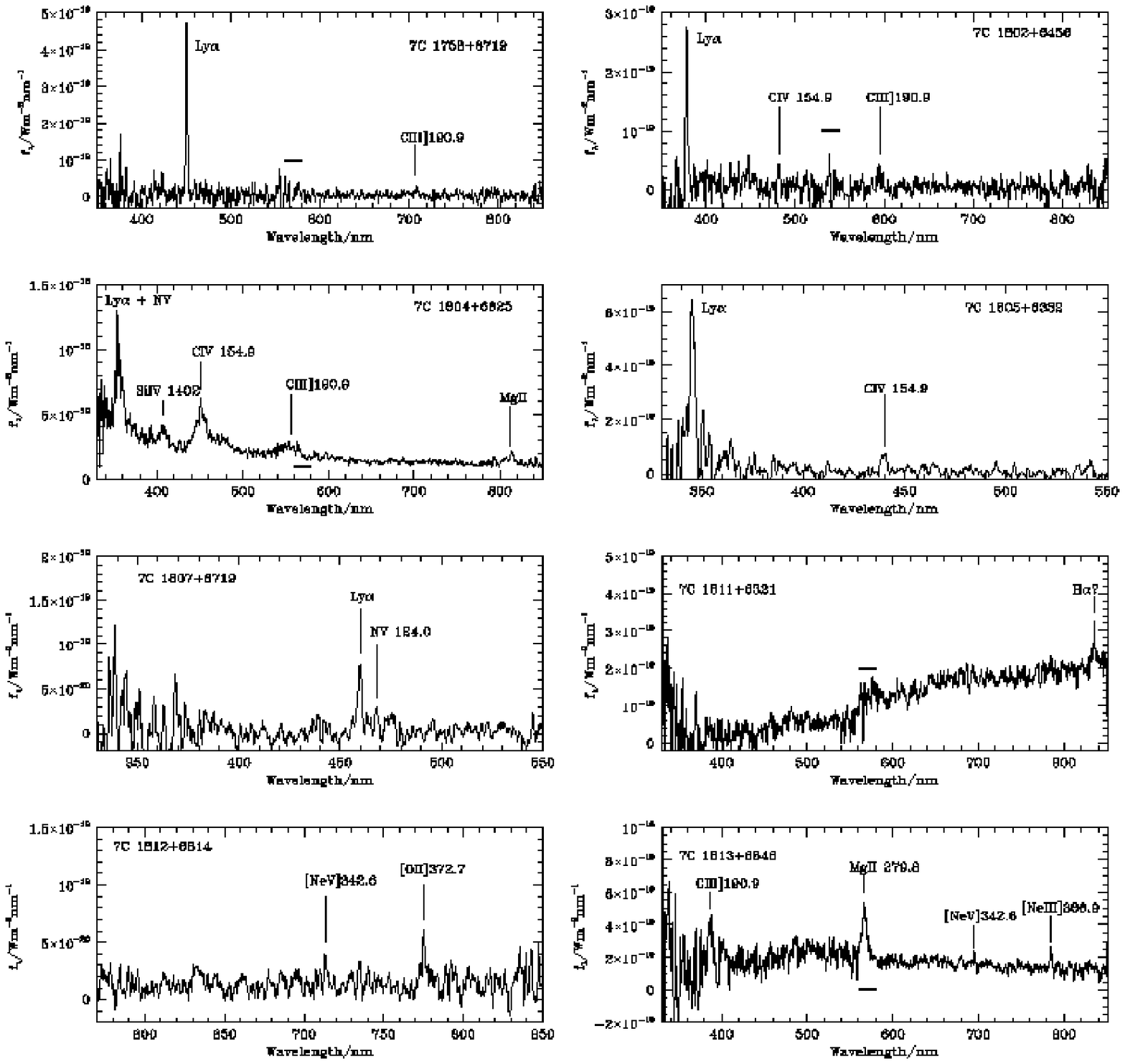}}
\end{picture}
\caption{ continued.}
\end{figure*}

\setcounter{figure}{0}

\begin{figure*}
\begin{picture}(400,660)
\put(-100,-60){\includegraphics{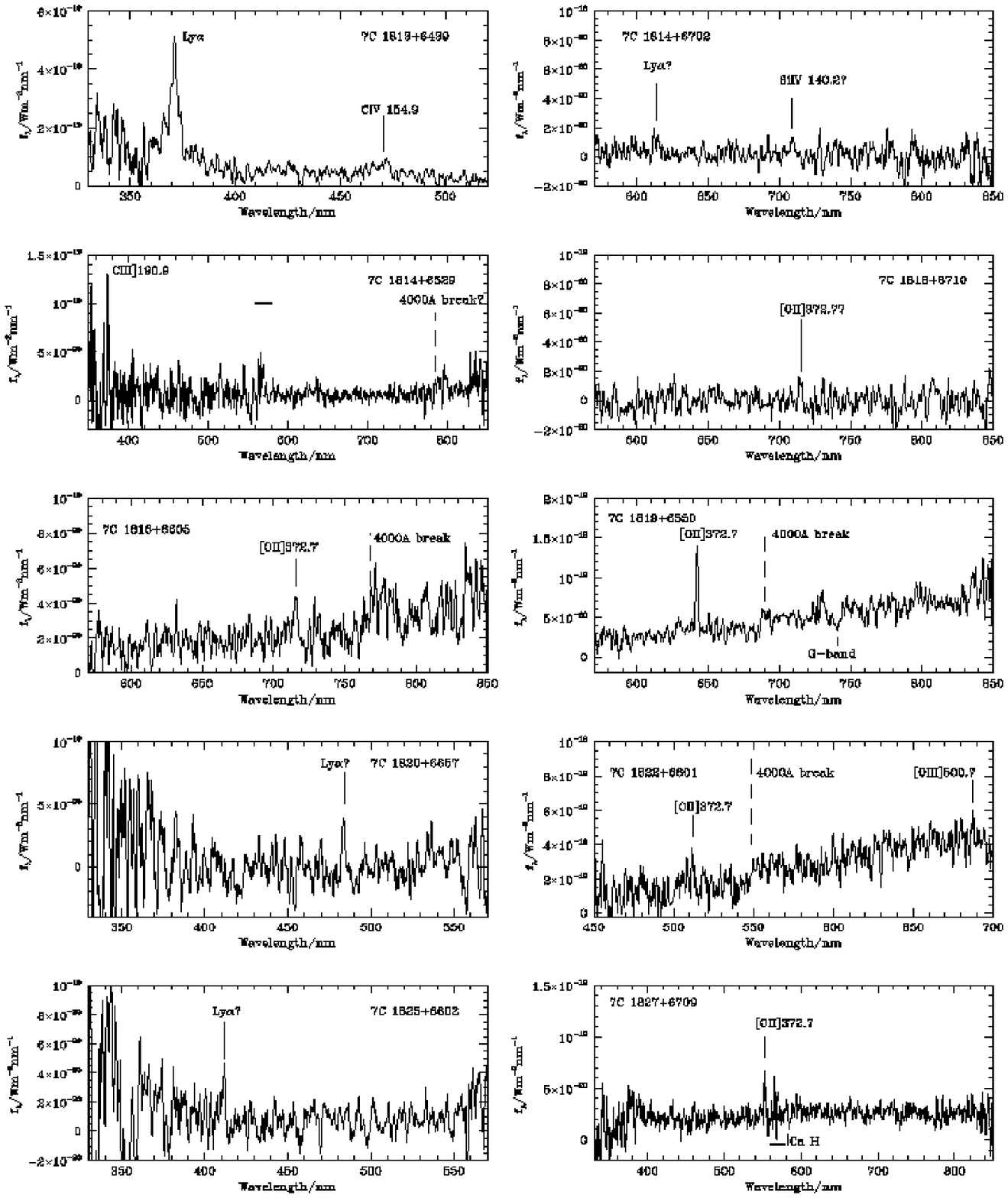}}
\end{picture}
\caption{continued.}
\end{figure*}

\begin{figure*}
\begin{picture}(200,120)
\put(-196,-355){\includegraphics{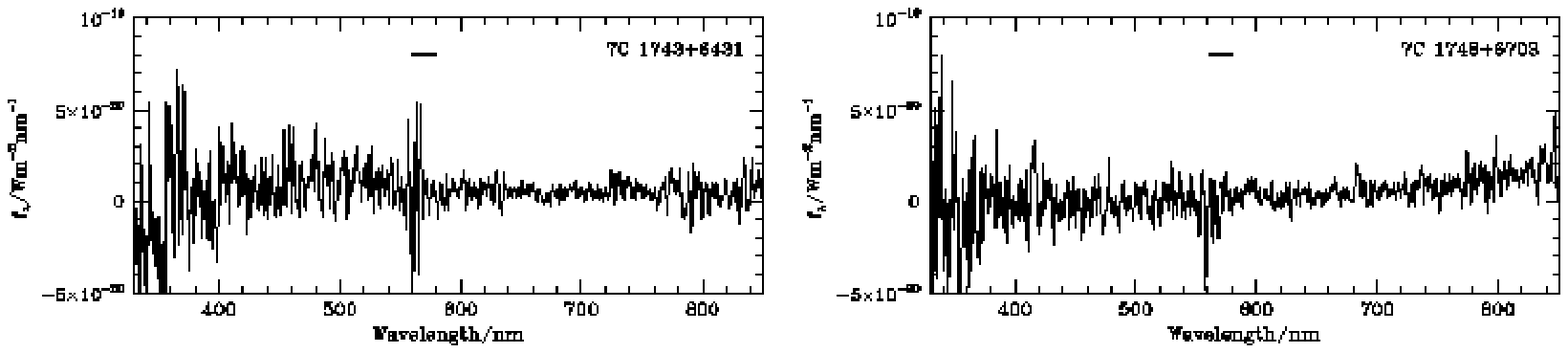}}
\end{picture}
\caption{Featureless spectra, displayed as in Fig.\ 1}
\end{figure*}

\subsection{Reliability of the redshifts}

The redshifts in Tables 4 and 6 come from spectra in which the signal:noise  
and number of spectral features used to estimate the redshift varied widely.
Those which we consider ``firm'' are based on more than one strong
spectral feature (emission or absorption lines or 4000${\rm \AA}$
break), and are graded $\alpha$. Those with either 
one strong spectral feature and one or more low signal:noise ones, 
or with two or more low signal:noise features 
are graded $\beta$. Those based on a single spectral feature
(for emission lines we usually assume to be Ly$\alpha$ if it appears in the 
blue arm spectrum or [O{\sc ii}]372.7 if it appears in the red), or two very 
weak features are graded $\gamma$, and should be considered 
unreliable until further data are obtained. 

For some objects we have 
only been able to place an upper bound on the redshift by detecting continuum
in the blue arm spectrum. Due to the low transmissivity of Lyman limit systems
we can be fairly sure that if, for example, we see continuum to the 
atmospheric 
cutoff (340 nm) then the redshift of the radio source is $< 340/91.2-1 = 2.7$

\subsection{Possible biases in the redshift distribution}

There are a number of observational problems which could lead to biases
in the measured redshift distribution of objects. The most 
serious bias probably comes from the difficulty of measuring redshifts
for objects with $1.2<z<1.8$, where [O{\sc ii}]372.7 moves out of the 
red arm spectrum, but before Ly$\alpha$ moves into the observable 
part of the blue. This is borne out by inspection of the redshift 
distribution which seems deficient in objects with $1.2<z<1.8$
compared to the predictions of the Dunlop \& Peacock (1990) models (Fig.\ 3).
Measurement of redshifts in this range depends on 
fainter features, typically Mg{\sc ii} 279.8 and C{\sc iii}]190.9; consequently
it is likely that many of the objects with no measured redshift
lie in this range. The radio luminosity function derived from the 7C 
redshift surveys will be discussed in a future paper (Willott et al.\ 1999).

\begin{figure}
\begin{picture}(100,240)
\put(-20,-80){\includegraphics{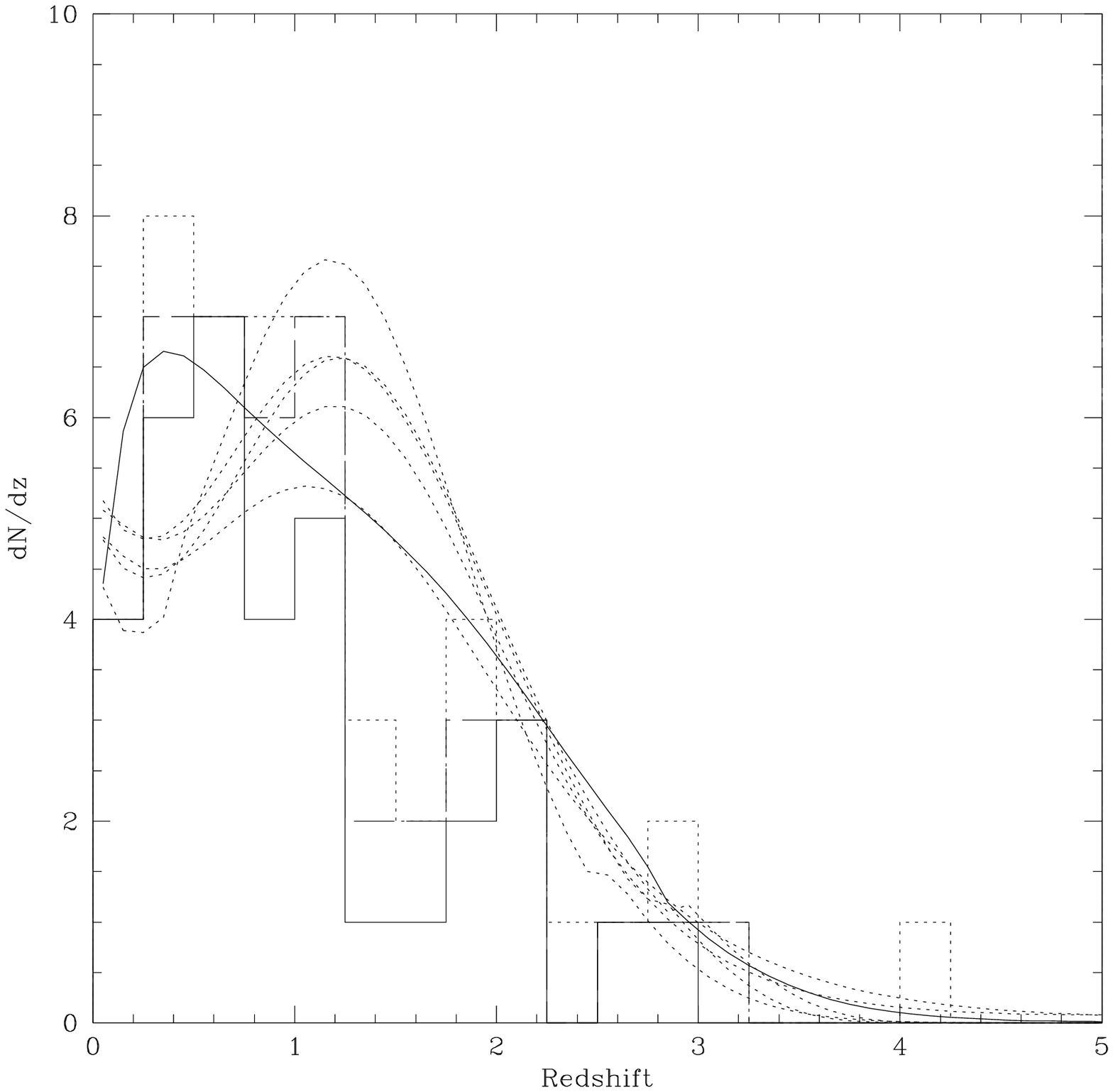}}
\end{picture}
\caption{Redshift distribution of the 7C-{\sc iii} sample. The distribution of 
grade $\alpha$ redshifts is shown as the solid histogram. $\beta$ redshifts 
are added in as dot-dash lines and $\gamma$ redshifts as dotted lines.
The curves are the predictions of the Dunlop \& Peacock (1990) models: 
the dotted lines are the free-form models and the solid line the 
luminosity-density evolution model.} 
\end{figure}

Another possible source of bias is the existence of objects 
with anomalously low Ly$\alpha$ emission (e.g.\ van Ojik et al.\ 1994, 
Dey et al.\ 1995). So far, all examples of 
these have fairly strong high ionisation lines, but this may simply be a 
selection effect, as objects without these would be very hard to obtain 
redshifts for. If Ly$\alpha$ extinction is common, then many of the 
objects without redshifts and without continuum emission 
in the blue arm could be very distant. 

Finally, it has recently become apparent that many of the highest redshift
radio sources are amplified by gravitational lensing effects. For 
example two well-studied $z>4$ radio galaxies have foreground galaxies within
a few arcseconds of the line of sight (Lacy et al.\ 1994; Rawlings et al.\ 
1996), and significant overdensities of foreground galaxies have been found
in the fields of high redshift 3C radio sources (Ben\'{\i}tez et al.\ 1997). 
Misidentification of the radio galaxy 
with a lensing object could lead to a substantial underestimate of the 
radio source redshift. Accurate astrometry is the best safeguard against 
this, but also lensing galaxies are unlikely to have the high 
equivalent width  emission lines frequently found in radio galaxies. We 
therefore feel that this is probably not important for our sample, where
nearly all the $z>1$ objects with established redshifts have spectra 
characteristic of active galaxies. We note, however, several cases in 
which a low redshift galaxy lies close enough to the radio structures
to be mistaken for the identification, but which we found through imaging or
spectroscopy were foreground objects. Some probable lensing events are 
discussed in Section 7.

%\setlength{\unitlength}{1mm}
%\begin{figure}
%\begin{picture}(75,75)
%\put(0,-20){\special{psfile=zdist.ps hscale=40 vscale=40}}
%\end{picture}
%\caption{Redshift distribution for the 7C-{\sc iii} sample}
%\end{figure}

\begin{table*}
\caption{Results of spectroscopy of the 7C-{\sc iii} radio sources}
\begin{tabular}{llcclrrrl}
%include spectrophotometry here...
Name & Redshift &Grade&Emission & Flux      & Width      &Equivalent& 
Comments \\
     &      &    & line     & /$10^{-20}$Wm$^{-2}$& /kms$^{-1}$&
width/nm  &     \\
7C 1731+6641& 0.561 & $\alpha$ & 
Mg{\sc ii}279.8&100&10500 & 19.2 & broad component \\
&                  &   &[O{\sc ii}]372.7&50&- &4.9 &  \\
&                  &   &[Ne{\sc iii}]386.9&26&- & -  &  \\
&                  &   &H$\beta$          &17&-&2.8&  \\
&                  &   &[O{\sc iii}]495.9 &30&-&4.9&  \\
&                  &   &[O{\sc iii}]500.7 &68&-&11.2&  \\     
7C 1732+6535* & 0.856 &$\alpha$?  & & & &   & Literature $z$ (Paper II) \\
%7C 1732+6715&  - &-& ? &       &&    -  &  missed\\  
7C 1733+6719*& 1.84 &$\beta$ & Ly$\alpha$  & 22   & - & -   &   \\  
       &     &      & Mg{\sc ii}279.8   &164&3000& 16.1&broad component \\
7C 1736+6710*& 0.188 &$\alpha$ &  &    & &    & Paper II \\   
        &    &       &  &    & &    &    \\
7C 1740+6640&2.10&$\alpha$ & Ly$\alpha$    &41  &-  &  -  & double \\  
            &  &  & C{\sc iv}154.9& 3  & - &  -  &      \\
7C 1741+6704*&1.054&$\alpha$&[O{\sc ii}]372.7& 29 & -& -  &  \\
            &     &&[Ne{\sc iii}]386.9&22&  -   & -  &  \\
7C 1742+6346*& 1.27&$\beta$  & Mg{\sc ii}279.8& 10 & - & - & \\
            & &   & [Ne{\sc v}]342.6& 8 & - & - & uncertain\\
            & &   & [O{\sc ii}]372.7& 8 & - & - & uncertain\\ 
7C 1743+6344*& 0.324 &$\alpha$& [O{\sc ii}]372.7 &8 & - & - &ID uncertain, see notes\\
            &   &     & [Ne{\sc iii}]386.9&4& - & - &\\
            &    &    & [O{\sc iii}]500.7 &5& - &- & \\
7C 1743+6431*&      ? &    -  &                   & && &  no lines\\
            &   &    &                      & & &  &          \\
7C 1743+6639*& 0.272 &$\alpha$    &                    & & & &Paper  II   \\
7C 1745+6415*& 0.673 &$\alpha$    &[O{\sc ii}]372.7 & 80  & - & 280 &LRWLR\\
& &    &H$\beta$          &  40             &     & 63  &  \\
& &    &[O{\sc iii}]495.9 & 140  &&170 &  \\
& &   &[O{\sc iii}]500.7 & 400  &&460 &  \\
7C 1745+6422*& 1.23 &$\alpha$&  &   & & &  K95\\    
7C 1745+6624& 3.01 &$\beta$ & Ly$\alpha$      & 17  &  &   &    \\
            &    &   & Si{\sc iv} 140.2& 5   &  &   & uncertain   \\
            &    &   & He{\sc ii} 164.0& 3   &  &   & uncertain   \\
7C 1747+6533*& 1.516&$\alpha$&  &   & &   &  K95   \\
            &   &      &                   & & &  &          \\
7C 1748+6703*& ?&-&    &    & &  &see notes  \\
7C 1748+6657*&1.045 &$\beta$&[O{\sc ii}]372.7&26&- &17&plus continuum break\\
7C 1748+6731*& 0.56 &$\alpha$ &[O{\sc iii}]500.7?&9 &- &1&LRWLR  \\   
7C 1751+6809*& 1.54 &$\beta$ &[Ne{\sc iv}]242.4 & 4  &- & -  & \\
            &  &   &Mg{\sc ii} 279.8  & 2  &- & -  & \\
            &  &    &                   & & &  &          \\
7C 1751+6455*& 0.294&$\alpha$ &&&& &absorption line $z$\\
7C 1753+6311*& 1.96 &$\gamma$&He{\sc ii}&7&-&-&+Ly$\alpha$ absorption?\\     
7C 1753+6543*& 0.140 &$\alpha$& & &  &   & R.G.\ McMahon (p.c.)    \\
7C 1754+6420*& 1.09&$\beta$& [O{\sc ii}]372.7  & 19& - & - & diffuse \\
            &   &   & [Ne{\sc v}]342.6 &  7& - & - &\\
            &   &    &                   & & &  &          \\
7C 1755+6314*& 0.388&$\alpha$ &      &   &&-&absorption line $z$\\   
7C 1755+6830*& 0.744&$\alpha$ &[O{\sc ii}]372.7& 20 &- &2.4&LRWLR\\
            &  &    & [Ne{\sc iii}]386.9&17&- &- &\\
            &  &    & [Ne{\sc v}]342.6?  &20&- & -&\\ 
7C 1756+6520*& 1.48&$\gamma$&[Ne{\sc iv}]242.4? & 4  &&  &see notes \\
7C 1758+6535*& 0.80&$\alpha$ &[O{\sc ii}]372.7 & 18   && 14 &LRWLR  \\
            &    &  & [Ne{\sc iii}]386.9& 10  & - & -  &    \\
7C 1758+6553*& 0.171 &$\alpha$ &   &   & &   &absorption line $z$  \\
            &   &    &                       & & &  &          \\
7C 1758+6307*& 1.19 &$\alpha$&C{\sc iii}]190.9&45  &5100 & -  &  \\
            &    &  &Mg{\sc ii} 279.8&41  &6600 &31  &  \\
            &    &  &[O{\sc ii}]372.7&12  &- &17  &  \\
7C 1758+6719*& 2.70 &$\alpha$& Ly$\alpha$ (north) & 120 & - & - & double\\
            &    &  &  Ly$\alpha$ (south)? &  30 & - & - & 
-38000 kms$^{-1}$ \\
            &    &  & C{\sc iii}]190.9   &   8 & - & - & from north 
component.\ \\
\end{tabular}
\end{table*}
\setcounter{table}{3}
\begin{table*}
\caption{Results of spectroscopy of the 7C-{\sc iii} radio sources continued}
\begin{tabular}{llclrrrl}
Name & Redshift &Grade&Emission & Flux      & Width      &Equivalent& 
Comments \\
     &         &      & line    & /$10^{-20}$Wm$^{-2}$& /kms$^{-1}$&
width/nm  &     \\
7C 1801+6902*& 1.27&$\alpha$ &           &    &   &    & Paper II \\
7C 1802+6456*& 2.11 &$\alpha$&Ly$\alpha$ & 71 & - & 48 &  \\
            &    &     &C{\sc iv} 154.9& 10& - & -  &  \\
            &    &     &C{\sc iii}]190.9& 8 & - & - &  \\
7C 1804+6625& 1.91 &$\alpha$ &Ly$\alpha$& 270& 3000 &5 &   \\
            &      &              &N{\sc v} 124.0&110&3700&2& \\
            &   &  &      Si{\sc iv} 140.2&120& 8000 &5 \\
            &   &  &      C{\sc iv} 154.9 &470& 11000&20 & \\
            &   &  &      He{\sc ii} 164.0& 160 & 11000&7&width fixed
to C{\sc iv} \\   
            &    &  &     Mg{\sc ii} 279.8&90& 4500 &8 \\
            &    &        &                   & & &  &          \\
7C 1804+6313*& ?    &- &       &         &   &       &  missed\\
7C 1805+6332*&1.84 &$\alpha$ & Ly$\alpha$     & 170   &1800 & &see notes\\
            &      &  &  C{\sc iv} 154.9&  25   &1800 & & \\
7C 1807+6831*& 0.58 &$\alpha$ & [O{\sc ii}]372.7& 11 & - & - &LRWLR\\
            &      & &  [Ne{\sc iii}]386.9& 8& - & - & \\
            &      & &  [O{\sc iii}]500.7& 22& - & - & \\            
7C 1807+6719*& 2.78 &$\alpha$ & Ly$\alpha$  & 21 & - & -  &  \\
            &      &   & N{\sc v} 124.0 &7& - & - &   \\ 
7C 1807+6841*& 0.816&$\alpha$ & [O{\sc ii}]372.7  & 52& 870 & 21&LRWLR\\
            &    &  &  [Ne{\sc iii}]386.9& 11& 580 &  4&  \\
            &    &      &                   & & &  &          \\
7C 1811+6321*&0.273&$\gamma$&H$\alpha$?&15& - &- &  \\
7C 1812+6814&1.08 &$\alpha$&[Ne{\sc v}]342.6& 5  & - & -  &  \\
            &    &  & [O{\sc ii}]372.7&10  & - & -  &  \\
7C 1813+6846*&1.03  &$\alpha$ &C{\sc iii}]190.9&150&- &10&  \\
            &      &  &Mg{\sc ii} 279.8&170& 3000 & 7 &\\
            &      &  &[Ne{\sc v}]342.6& 10&  -   & 0.6&\\
            &      &  &[Ne{\sc iii}]386.9&26& -   & 2  &\\
7C 1813+6439*&2.04  &$\alpha$&Ly$\alpha$&4100&200&20& \\
            &      &   &C{\sc iv} 154.9&6700&49&135&\\
7C 1814+6702*& 4.05&$\gamma$&Ly$\alpha$&5 &- &- &see notes\\
            &     &  &Si{\sc iv} 140.2&3&-&-&\\
7C 1814+6529*& 0.96&$\beta$&C{\sc iii}]190.9&31&-& -&plus 4000\AA$\;$break\\
            &     &       &                   & & &  &          \\
7C 1815+6805*& 0.230 &$\alpha$& -  & -  &- &    -  & Paper II; K95  \\
7C 1815+6815*& 0.794 &$\alpha$&Mg{\sc ii} 279.8&20&-&-&plus 4000\AA$\;$break; LRWLR\\
7C 1816+6710*& 0.92  &$\gamma$ & [O{\sc ii}]372.7& 4&-&-&see notes\\
7C 1816+6605*& 0.92  &$\beta$&[O{\sc ii}]372.7& 6&-&3&plus 4000\AA$\;$break\\
7C 1819+6550*& 0.724 &$\alpha$&[O{\sc ii}]372.7&20&-&6&plus 4000\AA$\;$break\\
            &      & &                       & & &  &          \\
7C 1820+6657& 2.98 &$\gamma$&Ly$\alpha$?&10& - & - &\\
%7C 1821+6419& 0.298&-   &                   & & &     & \\
7C 1822+6601*& 0.37  &$\alpha$ & [O{\sc ii}]372.7&40&-&2.4 & \\
            &     &  & [O{\sc iii}]500.7&12&-&0.3 & \\    
7C 1825+6602*& 2.38&$\gamma$ & Ly$\alpha$? & 7 & - &  - & see notes\\
7C 1826+6510*& 0.646   &$\alpha$ & -    & -   & - &  - &LRWLR \\
7C 1826+6704*& 0.287 &$\alpha$ & -   & -   & - &    -  & Paper II \\
            &       &   &                   & & &  &          \\
7C 1827+6709*& 0.48  &$\beta$ &[O{\sc ii}]372.7&10 & - & 5 &plus Ca H (Ca K
lost in sky line) \\
\end{tabular}

Notes: Linewidths are only given for emission lines significantly
broader than the instrumental width ($\approx$ 
1500 kms$^{-1}$ at 500 nm for the WHT data). Equivalent widths are
only given where the continuum is detected at high signal:noise. Asterisks  
by the names indicate members of the NEC* sample defined in Section 9.
The redshift grades are $\alpha$ for a firm redshift, $\beta$ for a less
certain redshift and $\gamma$ for an uncertain one, as discussed in 
Section 4.1. A dash indicates that no redshift could be measured.
\end{table*}

\subsection{Notes on individual objects}

\subsubsection*{7C 1740+6640}

This radio galaxy was observed in $R$-band through the Auxiliary Port
of the WHT on 1995 July 30 (Fig.\ 4). The airmass was 1.8, and the seeing
0.8-arcsec. The identification has $R=24.8$ in a 3-arcsec diameter aperture.

The Ly$\alpha$ line seen in the spectrum of 
this object appears to have a weak blueshifted 
component at $z=2.081$, compared to $z=2.109$ for the main Ly$\alpha$ 
component.

\begin{figure}
\begin{picture}(200,200)
\put(-185,-540){\includegraphics{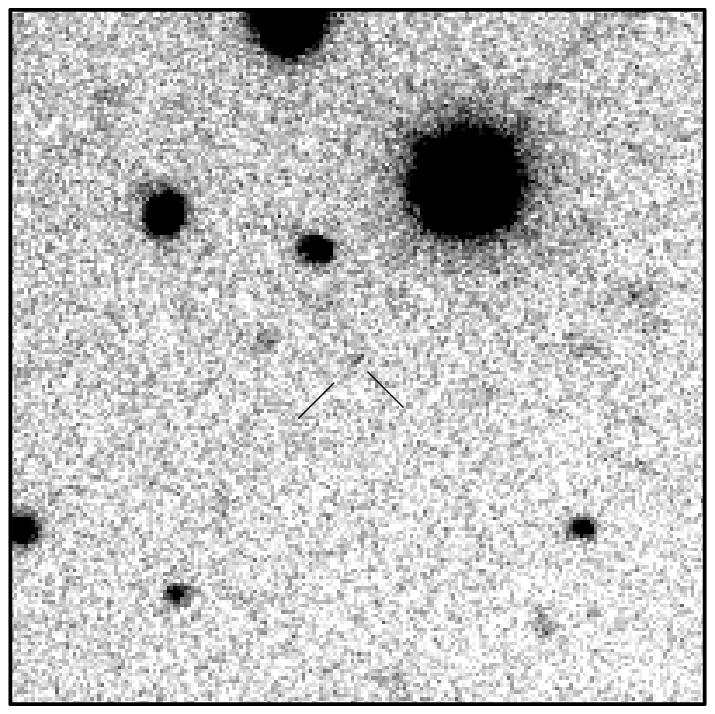}}
%\put(90,80){\line(1,1){10}}
%\put(110,93){\line(1,-1){10}}
\end{picture}
\caption{$R$-band image of 7C1740+6640. The image is 20 arcsec square,
and the position of the identification is marked. North is to the top and 
east to the left.}
\end{figure}

\subsubsection*{7C 1742+6346}

The redshift of this object is uncertain as both the 
[Ne{\sc v}] and [O{\sc ii}] emission line candidates are in regions of high 
sky noise, so may not be real. No emission was detected from the region of 
the putative central component. 
A number of other objects were present along the slit (see image in
Paper III). Galaxy `a' has emission lines at $z=0.326$, and may be a 
member of the nearby cluster A2280 whose cD galaxy is 4.5 arcmin distant and 
has $z=0.330$ (Paper II). Galaxy
`b' has a single emission line consistent with [O{\sc ii}]372.7 at $z=0.457$.

\subsubsection*{7C 1743+6344}

The identification of this object in Paper II is about 4-arcsec due
north of the mid-point of the radio hotspots. In our spectrum emission
lines at $z=0.324$ and very faint continuum are seen close to the
mid-point of the radio hotspots, and the original candidate
identification also appears to have [O{\sc ii}] emission at the same
redshift. Whichever of these two objects is the correct
identification, this radio source is very unusual in having such an apparently
subluminous optical counterpart. The redshift is consistent with the
radio source being a member of the A2280 cluster, whose cD galaxy is
nearby (Paper II).

\subsubsection*{7C 1743+6431}

The original identification of this radio
source in Paper II is probably incorrect. As discussed in 
Paper III, an object on the $H$-band
image is coincident with the position of a tentative radio central
component in Paper I. The $H$-band object is very faint in the optical
and no emission lines are discernible in the WHT spectrum, although
the detection of continuum down to 447nm allows us to constrain the
redshift to $<3.9$. The photometric redshift obtained assuming $H-K=1$ 
and the $K-z$ relation of Eales et al.\ (1997) is $\approx 1.7$, consistent
with no strong lines being seen in the spectrum and the very red $R-K$ colour. 
The original identification has a redshift of 
0.364, and is close to the western hotspot. It may therefore be
magnifying the hotspot through gravitational lensing effects.

\subsubsection*{7C 1745+6624}

This object was also observed on the NASA Infrared Telescope Facility 
(IRTF) on 
1998 February 15 for 30 min through the $K^{'}$ filter
(Fig.\ 5). The airmass was 1.7.
An object is just detected at the $3-\sigma$ level 0.6 arcsec
E and 1.1 arcsec south of the position of the radio point source, 
with $K\approx 20.8$ in a 2-arcsec diameter aperture. 

Although only the
Ly$\alpha$ line is strongly detected in the spectrum, the faint $K^{'}$ 
magnitude is consistent with the relatively high redshift of 3.01, as are 
marginal detections of Si{\sc iv} 140.2 (unfortunately in the dichroic region)
and He{\sc ii} 164.0, along with a possible continuum break across
L$\alpha$.

\begin{figure}
\begin{picture}(200,240)
\put(-120,-220){\includegraphics{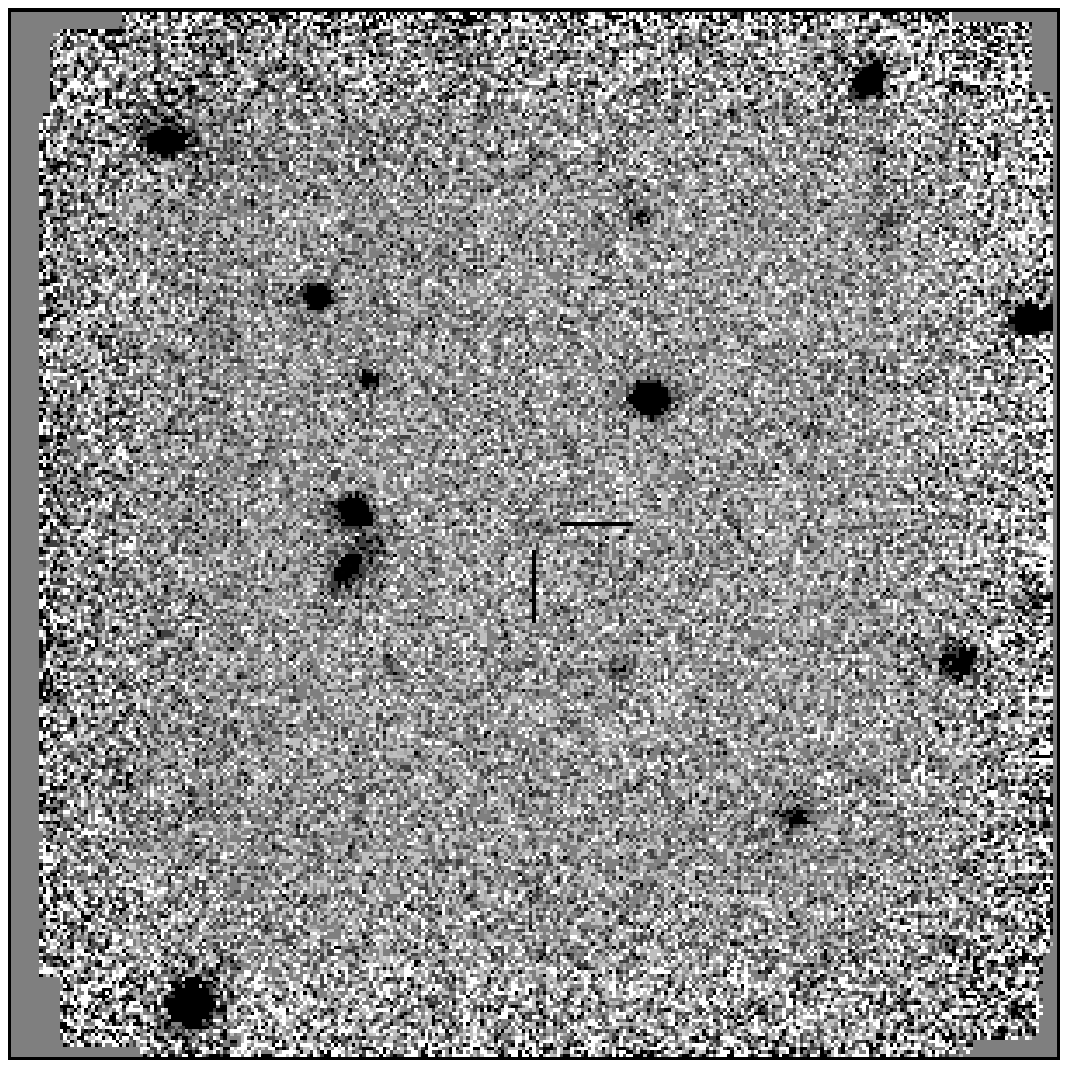}}
%\put(115,90){\line(0,1){15}}
%\put(120,110){\line(1,0){15}}
\end{picture}
\caption{$K^{'}$ image of 7C1745+6624. The image is 93 arcsec square,
and the position of the identification is marked. North is to the top and 
east to the left.}
\end{figure}

\subsubsection*{7C 1748+6703}

No emission is seen from a spectrum centred on the radio position of
the assumed radio central component. In Paper III we discuss the 
detection of a galaxy $\approx 2$ arcsec north of this
position, in addition to a very faint object coincident with the
radio central component. The galaxy $\approx 10$ arcsec east of the radio
position has an emission line at 683.5 nm and a red continuum. If the
emission line is [O{\sc ii}] the redshift of this galaxy is $0.83$. 
Both this galaxy and the galaxy to the north are probably
close enough to the radio source to be magnifying it via gravitational
lensing if the redshift of the radio galaxy is $\gg 1$.

\subsubsection*{7C 1753+6311}

This object has a very unusual spectrum, with very blue continuum
colours (approximately flat in $f_{\nu}$), 
but no strong emission lines. We have tentatively 
identified an apparent
break in the UV spectrum with Ly$\alpha$ absorption at $z=1.96$,
also consistent with a weak emission feature at 485 nm being 
He{\sc ii} 164.0. The apparent break may, however, be due to a
combination of the atmospheric cutoff and low signal-to-noise, so the
redshift must be considered tentative until at better spectrum can be
obtained. The EW of the UV emission lines must be fairly low for us not
to have detected any. Indeed 
the spectrum is much more reminiscent of starburst galaxies than of
AGN, suggesting a starburst may be the origin of the blue light; in
this case the AGN must be either very weak compared to some of our other 
$z\sim 2$ radio galaxies or obscured by the starburst. This source
also has an unusual radio structure (Paper I).

\subsubsection*{7C 1756+6520}

The spectrum of this object shows very faint ($R\approx 24$) 
continuum throughout the spectrum
which is approximately constant in $f_{\lambda}$. There is a
marginally-significant emission line at 600.5 nm. It is unlikely to be
Ly$\alpha$ due to the lack of any continuum drop shortward of the
emission line wavelength, and in particular the presence of flux below
the redshifted Lyman limit at 450 nm. On the other hand, the faintness
of the continuum emission suggests a higher redshift than the 0.6
derived from assuming the line is [O{\sc ii}]372.7. We have therefore
assigned a very tentative redshift of 1.48 to this object, assuming
that the emission line is [Ne{\sc iv}]242.4, one of the brightest
lines for which we would not expect to
see bright Ly$\alpha$ or [O{\sc ii}]372.7 emission in the
spectrum. In the
composite spectrum of McCarthy (1994), this and C{\sc ii}]232.6 are
comparably bright, but if we assume the line is [Ne{\sc iv}] then the 
C{\sc ii}] line would be in the dichroic region and therefore quite
likely to have been missed.

\subsubsection*{7C 1758+6719}

This object was originally identified with a faint optical object 
approximately midway between two radio components, which are
45 arcsec apart (Paper II).
We placed our slit through both radio components and the optical object.
Ly$\alpha$ and C{\sc iv} emission were seen from the northern component at 
$z=2.70$. We also found a single emission line from the approximate position
of the southern radio component at 395.2 nm, which, if Ly$\alpha$, would 
correspond to a redshift of 2.25. The original identification has a single
emission line at 665.3 nm, with a possible second line at 776.4 nm in the 
sky lines, which could correspond to [O{\sc ii}]372.7 and H$\gamma$ at 
$z=0.785$. HST, MERLIN and VLA imaging and together with further optical 
spectroscopy of this enigmatic radio source will be presented in a future 
paper.

\subsubsection*{7C 1802+6456} 

A galaxy with strong emission lines at $z=0.69$ is seen within
2-arcsec of the
western hotspot. It is probably magnifying the hotspot emission via
gravitational lensing.

\subsubsection*{7C1804+6313}

This object has no clear identification. A marginally-resolved radio
component was discovered in Paper I, but subsequent VLA mapping presented in 
Paper III showed that this was just one end of a low surface brightness 
radio galaxy. A spectrum was taken through objects `a', `b' and `F'
of Paper III, and a further one through `a' alone;  all these objects
seem to be stars. The most likely identification is therefore
probably object `c' of Paper III for which we have yet to obtain a spectrum.

\subsubsection*{7C 1805+6332}

The original identification of this object in Paper II
was an $R=21.1$ galaxy 3 arcsec from the line joining the midpoint 
of the radio hotspots. The spectroscopic observations were 
made with the slit placed
so as to cover both the proposed identification and the midpoint of the 
hotspots. The resulting spectrum shows that the true identification 
lies close to the midpoint of the hotspots, with the original 
identification  being 
foreground, with a probable redshift (based on H$\alpha$ in emission
and Mg{\sc i} in absorption) of 0.32. 
This is another example of an object which may be being 
amplified by gravitational lensing.

\subsubsection*{7C 1811+6321}

This object has a largely featureless red continuum, apart from one
strong emission line which is taken to be H$\alpha$. The nearby,
unaligned companion galaxy has no spectral features, but has a similar
spectral energy distribution to the radio galaxy.

\subsubsection*{7C 1814+6702}

This object shows very faint continuum between 615 and 700 nm. There
are two very marginally-detected emission line candidates whose ratio
of wavelengths is consistent with Ly$\alpha$:Si{\sc iv} 140.2 at
$z=4.05$ or [Ne{\sc iv}]242.4: Mg{\sc ii} at $z=1.53$. 
As there is no sign of continuum emission shortward of the shortest wavelength
emission line candidate we have provisionally taken the redshift to be
$z=4.05$ pending further spectroscopy, although this should probably
be considered an upper limit on the true value of the redshift.

\subsubsection*{7C 1816+6710}

This object shows only a single faint, spatially extended emission line and 
faint continuum. We assume the line is [O{\sc ii}] as there is no sign
of a drop in the continuum shortward of the line wavelength.

\subsubsection*{7C 1825+6602}

Another object showing only a single, marginally-detected 
faint emission line and faint
continuum. The continuum emission is approximately constant in
$f_{\lambda}$ throughout both arms of the spectrum. We have
assumed the single emission line detected in the blue to be 
Ly$\alpha$ at $z=2.38$, but it is also possible that this object lies in
the $z=1.3-1.8$ range where there are no strong emission lines. The apparent 
blue continuum excess shortward of our assumed Ly$\alpha$ line is probably 
light refracted into the slit from a nearby star.

\begin{figure*}
\begin{picture}(400,400)
\put(-100,-250){\includegraphics{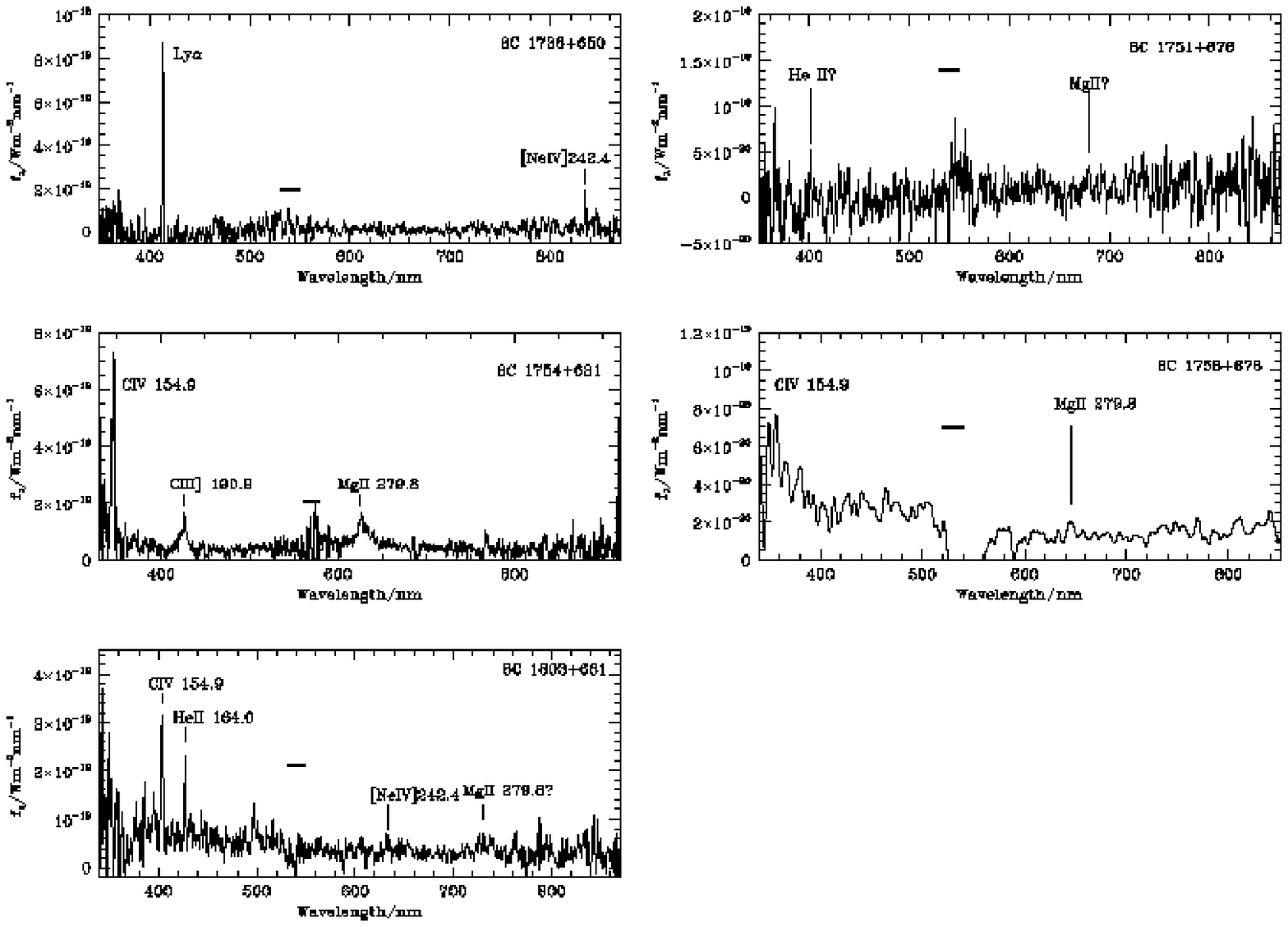}}
\end{picture}
\caption{Spectra of 8C-NEC objects not in the 7C-{\sc iii} sample}
\end{figure*}

\begin{table*}
\caption{Log of new spectroscopy of the 8C-NEC radio sources}
\begin{tabular}{lcclrcrr}
8C Name & Observing date(s)&Telescope & wavelength range(s) & integration
& slit width & slit PA& airmass\\
     &               &          &                 &  time(s)/sec & /arcseconds
 & \\
8C 1736+650&19/06/93&WHT&330 - 850 nm&1200&3&70&1.25\\
8C 1751+676&19/06/93&WHT&330 - 850 nm&3600&2&18&1.29\\
8C 1754+631&20/08/93&WHT&330 - 900 nm&1800&3&90&1.60\\
8C 1758+676&04/07/97&Shane&330 - 1060 nm&3600&2&50&1.3\\
8C 1803+661&17/06/93&WHT&330 - 850 nm&1200&3&122&1.26\\
\end{tabular}
\end{table*}

\begin{table*}
\caption{Results of spectroscopy of the 8C-NEC radio sources}
\begin{tabular}{lllclrrrl}
%include spectrophotometry here...
8C Name& 7C Name & $z$ &Grade &Emission & Flux      & Width      &Equivalent& 
Comments \\
&(if in 7C-{\sc iii} sample) & & & line     & /$10^{-20}$Wm$^{-2}$& /kms$^{-1}$&
width/nm  &     \\
8C 1732+655* & 7C 1732+6535& 0.856&$\alpha$?&   &    && &    Paper II    \\
%8C 1732+672 & 7C 1732+6715&  -   &-   &   &    &&-& new member (L98)\\  
8C 1733+673*& 7C 1733+6719& 1.84 &$\beta$ & & & - & -&new member (L98)\\  
8C 1736+650*&      -      & 2.40 &$\alpha$  &Ly$\alpha$&180&-&-&\\
            &             &      &        &[Ne{\sc iv}]242.4&23&-&-&\\
8C 1736+670*& 7C 1736+6710& 0.188 &$\alpha$ &  &    & &    & Paper II \\   
   &     &    &        &  &    & &    &    \\
8C 1741+670*& 7C 1741+6704&1.054&$\alpha$& && -& -  &  \\
8C 1742+637*& 7C 1742+6346&1.27 &$\beta$ & && -& -  &  \\
8C 1743+637*& 7C 1743+6344& 0.324 &$\alpha$&& & - & - &ID uncertain,\\
8C 1743+645*& 7C 1743+6431& ?     &   -     && &  -&  - & \\
8C 1743+666*& 7C 1743+6639& 0.272 &$\alpha$  &     & & & &Paper  II   \\
   &     &    &       &  &  &    & &        \\
8C 1745+642*& 7C 1745+6415& 0.673 &$\alpha$    &&   & - & &  \\
8C 1745+644*& 7C 1745+6422& 1.23 &$\alpha$ &  &   & & &  K95\\    
8C 1747+655*& 7C 1747+6533& 1.516&$\alpha$&  &   & &   &  K95   \\
8C 1748+670*& 7C 1748+6703& ?&-&  &    &    &  &see notes  \\
8C 1748+675*& 7C 1748+6731& 0.56 &$\alpha$ & &   &  & &  \\   
   &     &    &       &  &  &    & &        \\
8C 1751+681A*& 7C 1751+6809& 1.54 &$\beta$ &&  &- & -  & \\
8C 1751+681C  &     -     & ?    &         &&  &  &    &no spectrum\\
8C 1751+676*  &     -     & 1.42 &$\gamma$&He{\sc ii}164.0?&8&&&\\
              &           &      &         &Mg{\sc ii}279.8?&9&&&\\
8C 1751+649*  & 7C 1751+6455& 0.294&$\alpha$ &&&& &\\
8C 1753+631*  & 7C 1753+6311& 1.96 &$\beta$ &&&-&-&\\     
   &     &    &       &  &  &    & &     \\
8C 1753+664   &      -      & ?    &       &&&&&no spectrum\\ 
8C 1753+657*  & 7C 1753+6543& 0.140 &$\alpha$& &&&& \\
8C 1754+643*  & 7C 1754+6420& 1.09&$\beta$& & & - & - &  \\
8C 1754+631   &      -      &1.23 &$\alpha$&C{\sc iv} 154.9&220&3000&25&\\
              &             &     &         &C{\sc iii}]190.9&86 &6100&24&\\
              &             &     &         &Mg{\sc ii} 279.8&130&7300&21&\\
8C 1755+632*  &7C 1755+6314 & 0.388&$\alpha$ & & &&&\\   
              &             &     &          & & &&&\\
8C 1755+685B* & 7C 1755+6830& 0.744&$\alpha$ && & &&\\
8C 1757+653*  & 7C 1756+6520& 1.48&$\gamma$ & & &&  &\\
8C 1758+676   &      -      & 1.31&$\beta$  & C{\sc iv} 154.9&55&-&18&\\
              &             &     &          & Mg{\sc ii}279.8&6& - & 5 & \\
8C 1758+655*  & 7C 1758+6535& 0.80&$\alpha$ &&   &&  &    \\
8C 1758+658*  & 7C 1758+6553& 0.171 &$\alpha$ &   &   & &&  \\
   &     &    &       &  &  &    & &        \\
8C 1758+631*  & 7C 1758+6307& 1.19 &$\alpha$&&  & & &  \\
8C 1758+673*  & 7C 1758+6719& 2.70 &$\alpha$&&  & - & - &\\
\end{tabular}
\end{table*}
\setcounter{table}{5}
\begin{table*}
\caption{Results of spectroscopy of the 8C radio sources continued}
\begin{tabular}{lllclrrrl}
8C Name& 7C Name & $z$ &Grade& Emission & Flux      & Width      &Equivalent& 
Comments \\
&(if in 7C-{\sc iii} sample) & & & line     & /$10^{-20}$Wm$^{-2}$& /kms$^{-1}$&
width/nm  &     \\
8C 1801+690*& 7C 1801+6902& 1.27&$\alpha$&  & & &  &  \\
8C 1802+649*& 7C 1802+6456& 2.11 &$\alpha$ & & & & &  \\
8C 1803+661*&      -      & 1.61 &$\alpha$ & C{\sc iv} 154.9&53&-&6&\\
            &             &      &         & He{\sc ii} 164.0&26&-&4&\\
            &             &      &         & C{\sc iii}]190.9&19&-&3&\\
            &             &      &         & [Ne{\sc iv}]242.4&5&-&1&\\
            &             &  & &  Mg{\sc ii} 289.8?&49&9000&19&broad? \\
8C 1804+664 & 7C 1804+6625& 1.91 &$\alpha$   &&& & &   \\
8C 1804+632* & 7C 1804+6313& ? &- &       &         &   &       & \\
   &     &    &       &  &  &     &    &    \\
8C 1805+635*& 7C 1805+6332& 1.84 &$\alpha$   & &    & & &\\
8C 1807+685*& 7C 1807+6831& 0.58 &$\alpha$   && & - & - & \\
8C 1807+687*& 7C 1807+6841& 0.816&$\alpha$ &  && &&  \\
8C 1808+677 &      -      &   ?  &   -     &  & & & &no spectrum\\
8C 1811+633*& 7C 1811+6321&0.273&$\gamma$ &&& & &  \\
   &     &    &       &  &  &    & &        \\
8C 1813+687*&7C 1813+6846&1.03  &$\alpha$  &&& &&  \\ 
8C 1813+646*&7C 1813+6439&2.04  &$\alpha$  &&& && \\
8C 1814+670*&7C 1814+6702& 0.95&$\gamma$ & & & & &\\
8C 1814+657 &      -     & 0.95&$\beta$  & & & & & Paper II\\
8C 1814+655*&7C 1814+6529& 0.96&$\beta$&&&& &\\
          &  &     &  &      &                  & &  &          \\
8C 1815+680*& 7C 1815+6805& 0.230 &$\alpha$ & &  & &  & \\
8C 1815+682*& 7C 1815+6815& 0.794 &$\alpha$&&&&&\\
8C 1816+671*& 7C 1816+6710& 0.92  &$\gamma$ & & &&&\\
8C 1816+660*& 7C 1816+6605& 0.92  &$\beta$&& &&&\\
8C 1819+658*& 7C 1819+6550& 0.724 &$\alpha$&  &&&&\\
&            &      & &      &                  & &  &          \\
8C 1821+643A*&     -      & 0.298 &$\alpha$       &&&&&\\
8C 1821+643B&      -      &0.304 &$\alpha$?& & & & &Literature $z$ \\ 
8C 1823+660*& 7C 1822+6601& 0.37  &$\alpha$ & &&& & \\
8C 1826+660A*& 7C 1825+6602& 2.38&$\gamma$ & &  &  &  &\\
8C 1826+651*& 7C 1826+6510& 0.63   &$\alpha$ &  &   & &  & \\
8C 1826+670*& 7C 1826+6704& 0.287 &$\alpha$ &  &  & &  & \\
8C 1827+671*& 7C 1827+6709& 0.48  &$\beta$ && & & & \\
\end{tabular}

Notes: Linewidths are only given for emission lines significantly
broader than the instrumental width ($\approx$ 
1500 kms$^{-1}$ at 500 nm for the WHT data). Equivalent widths are
only given where the continuum is detected at high signal:noise. Asterisks 
by the names indicate members of the NEC* sample defined in Section 9.
The redshift grades are $\alpha$ for a firm redshift, $\beta$ for a less
certain redshift and $\gamma$ for an uncertain one, as discussed in 
Section 4.1. A dash indicates that no redshift could be measured.

\end{table*}

\section{Gravitational lensing}

A number of coincidences of foreground galaxies and clusters with 7C-{\sc iii}
and 8C-NEC sources have been noted in both this paper and Paper III. The best 
candidates for lensing are listed in Table 7, along with source and lens 
redshifts where known. There may also be others which will 
only become apparent once deeper images or spectra are obtained. High 
redshift radio sources present a large geometrical
cross-section to lensing, particularly when selected at low frequency where
the extended heads and lobes can still contribute significantly to the 
total flux. Because of this large size, however, magnification factors 
are typically modest, $\stackrel{<}{_{\sim}}2$ and the effect on the 
luminosity function may be fairly unimportant, at least at 
$z\stackrel{<}{_{\sim}} 4$ (Lacy 1998). Follow-up observations of these 
objects, together with the lensing statistics of the samples and their 
implications will be discussed in a future paper.

Studies of larger samples of high redshift radio sources 
should allow us to investigate the effect of 
lensing on the observed radio luminosity function at high redshift. 
We will also be able to make determinations of mass:light ratios for 
significant numbers of lensing galaxies in the range
$0.3 \stackrel{<}{_{\sim}} z \stackrel{<}{_{\sim}} 1$.

\begin{table}
\caption{Candidate gravitationally-lensed radio galaxies}
\begin{tabular}{lcll}
Name & $z_{\rm Source}$ & $z_{\rm Lens}$ & Notes \\ 
8C1742+637& 1.27        & 0.33?          & Weak lensing of both hotspots?\\
8C1743+645& $\sim $1.7? & 0.36           & Probable lensing of W hotspot\\
8C1802+649& 2.11        & 0.69           & Probable lensing of W hotspot\\
8C1805+635& 1.84        & 0.32           & Lobe is probably lensed\\
\end{tabular}

\end{table}

\section{Cosmic evolution of the radio source population}

As we have complete samples at both 38 and 151 MHz we are 
able to introduce an important refinement to the study of the cosmic 
evolution of radio sources. All previous studies have been limited to samples 
selected at a particular observed frequency, which corresponds to different 
rest-frame frequencies depending on the redshift of the radio source. Thus a 
sample selected at, say, an observed frequency of 151 MHz is effectively 
selecting $z=2$ sources at 453 MHz. Furthermore, all such samples are 
selected on the basis of flux density ($S_{\nu}$) rather than emitted 
flux.

Investigating the importance of this effect is hard, particularly at low radio
frequencies as the $K$-corrections are uncertain and usually have to rely on 
an extrapolation of the spectrum measured at high frequencies. In the NEC, we 
have the 38MHz fluxes to supply the low frequency
point for the 151MHz 7C-{\sc iii} sample, and a variety of deep, higher 
frequency radio surveys to derive an 
accurate radio source spectrum from. The surveys used are listed in Table 8.
Note that the comparable or higher angular resolution offered by many of 
these surveys means that confusion of source fluxes can be practically 
eliminated. Copies of the Table containing the flux information for the 
sources listed in this paper will be made available over the Internet 
(www-astro.physics.ox.ac.uk/$\sim$mdl).

\begin{table}
\caption{Surveys used to determine flux densities in the NEC}
\begin{tabular}{lll}
Frequency & Beamsize & Reference\\
38 MHz    & $4.5\times 4.5\;$cosec$\; \delta$ arcmin$^2$ & Rees 1990\\
151 MHz   & $70 \times 70\;$cosec$\; \delta$ arcsec$^2$  & Lacy et al.\ 1995\\
327 MHz   & $54 \times 54\;$cosec$\; \delta$ arcsec$^2$&Rengelink et al.\ 1997\\
408 MHz   & $3.4\times 3.4\;$cosec$\; \delta$ arcmin$^2$ &Roger, priv.\ 
comm.\ \\
1.5 GHz   & $\approx 20 \times 20\;$arcsec$^2$ &Kollgaard et al.\ 1994 \\
2.695 GHz &$4.5 \times 4.5\;$arcmin$^2$&Loiseau et al.\ 1988 \\
5 GHz      &variable&Paper I or Greenbank\\
\end{tabular}

\vspace*{0.1in}

\noindent
Note: Greenbank fluxes are from Gregory \& Condon (1991), or Becker, White 
\& Edwards (1991). 
\end{table}

\subsection{Selection of a new complete sample}

We have elected to select our sample based on $\nu S_{\nu}$ in the 
rest-frame of the radio source.
If we select with a flux limit of $(\nu_{\rm s} S_{\nu_{\rm s}})_{\rm Limit}$
at a selection frequency $\nu_{\rm s}$ in the rest-frame of the radio source, 
this is equivalent to a limit of 
\[ \frac{\nu_{\rm s}}{(1+z)} S_{[\nu_{\rm s}/(1+z)]} \geq (\nu_{\rm s} \
S_{\nu_{\rm s}})_{\rm Limit} \]
in the observed frame. We work out $S_{[ \nu_{\rm s}/(1+z) ]}$ using a fit to 
the radio spectrum described below. Cancelling out $\nu_{\rm s}$ our 
selection then reduces to imposing a redshift-dependent flux density limit:
\[\frac{S_{[ \nu_{\rm s}/(1+z)]}}{1+z} \geq (S_{\nu_{\rm s}})_{\rm Limit}. \]

We argue that this is a more 
physically useful way of defining a sample as the selection is based directly 
on power emitted at low frequency rather than on flux received per unit 
frequency. There are of course many alternative selection methods one 
could think of (for example
integrating the spectrum in the rest frame over some larger frequency range), 
but this criterion has the advantage of being simple to implement.
For the NEC samples we have used a  selection criterion of 
$S_{151 ({\rm rest})}/(1+z)\geq 0.45$ Jy, obtaining $S_{151 ({\rm rest})}$ 
from fitting quadratics in log frequency versus log flux density to obtain 
smooth spectra in the range 38 - 4850 MHz. The flux density limit of the 
8C-NEC sample is 
$S_{38}=1.3$ Jy in the observed frame, comfortably below the limit for a 
$z\approx 3$ source 
from the 7C-{\sc iii} sample which has its rest-frame 
151-MHz flux emitted at 38 MHz, namely $S_{38}=2$Jy. Objects which are bright
enough to meet the selection criteria for this sample, which we shall refer to 
as the NEC* sample, are marked with an asterisk in Tables 4 and 6.  
Rest-frame 151-MHz flux densities and 1-GHz spectral indices are given in 
Table A2 of the Appendix. 

For sources
without redshifts, if they lie above the flux limits of the sample in the
observed frame, 0.5 Jy at 151 MHz and 2Jy at 38 MHz, they are counted as being 
in the NEC* sample, and conversely if they lie below both flux limits they
are not. This test is ambiguous for only one object, 8C 1808+677. This object 
has $S_{38}=2.0$ Jy and will therefore only be included if it is at $z\geq 3$; 
we therefore have excluded this object from the sample. 

The resulting sample contains 51 objects, 33 of which have $\alpha$ redshifts, 
eight $\beta$ redshifts, seven $\gamma$ redshifts and three objects with no 
redshift information.

The primary effect of the $\nu S_{\nu}$ selection is to remove the 
compact steep spectrum (CSS), objects from the 7C-{\sc iii} sample.
These drop out as they often have spectra which are flattening
towards low frequency. They are replaced with extended, steep 
spectrum objects \footnote{
Absorption of low frequency emission in our Galaxy is 
unlikely to be causing objects to drop out through spectral flattening at 
low frequency. Kassim (1989) finds that supernova remnants at galactic 
latitudes $|b| \stackrel{>}{_{\sim}} 3^{\circ}$ have no evidence for low 
frequency turnovers at 30.9 MHz, although these are common for 
remnants closer to the plane and seen towards the inner galaxy. For comparison 
the NEC is at galactic coordinates $l^{\rm II}=96^{\circ}, 
b^{\rm II}=+30^{\circ}$, so on 
that basis galactic absorption is unlikely to be significant.
Perhaps the best argument against galactic absorption, however, is that 
the objects which do show spectral flattening at low frequencies tend to be
CSS sources, where there are plausible physical reasons for an intrinsic 
turnover in the source spectrum.}.

The redshift distributions of the NEC*, 8C-NEC and 7C-{\sc iii} 
samples are compared in Fig.\ 7. As can be seen there are no strong 
dependencies of the redshift distribution on the sample selection method. 
The largest change is seen at $z>2$, where three of the 7C-{\sc iii} 
high-$z$ CSS 
sources drop out of both the NEC* and the 8C-NEC samples, and one extended
$z>2$ object (8C1736+650) comes in to both the NEC* and 8C-NEC samples.

\begin{figure}
\begin{picture}(100,240)
\put(-20,-80){\includegraphics{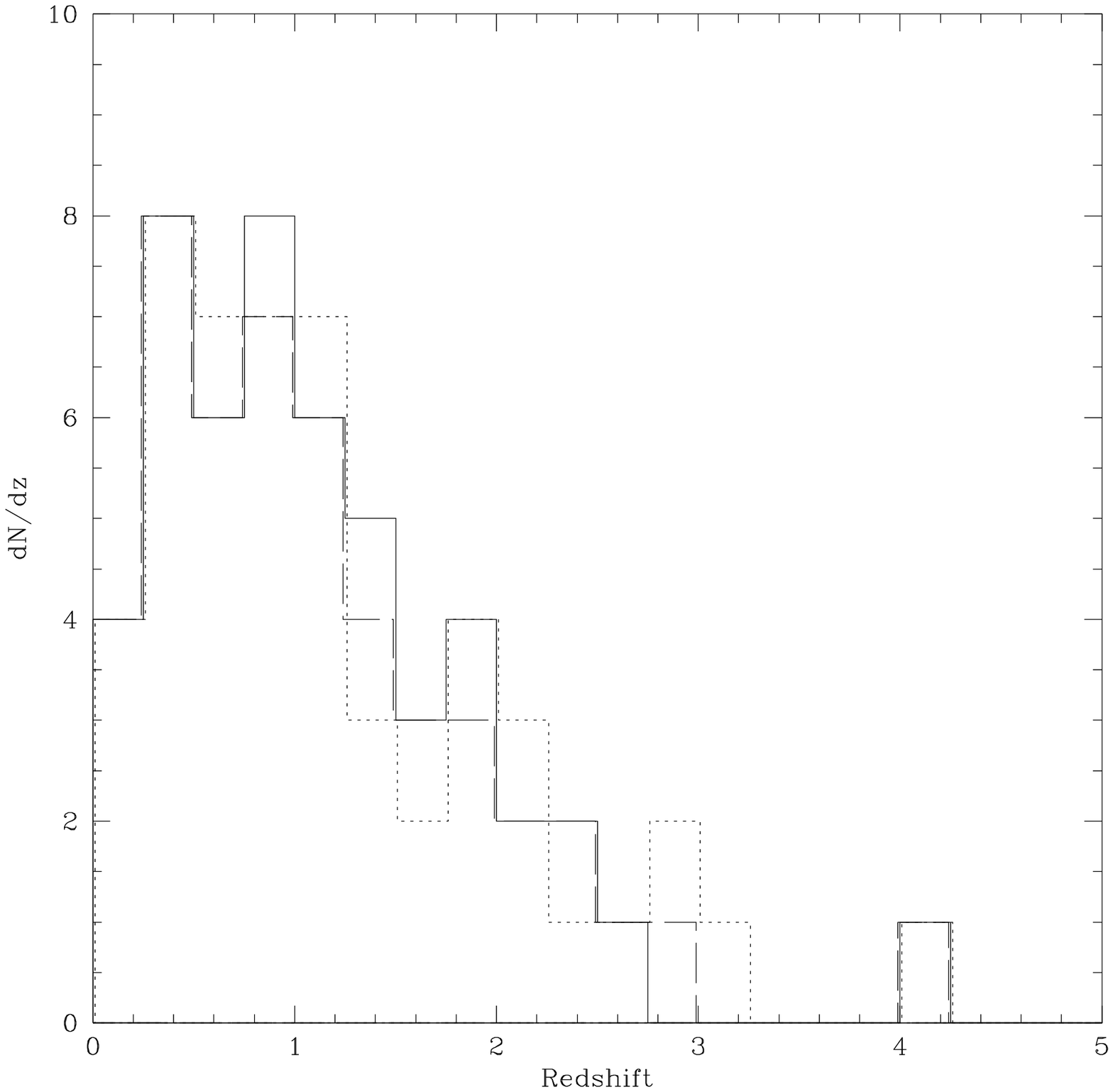}}
\end{picture}
\caption{The comparative redshift distributions of the 8C-NEC, 7C-{\sc iii} and
NEC* samples. The 8C-NEC sample is the solid histogram, the short-dash the 7C-{\sc iii} and the long-dash the NEC*.}
\end{figure}

For comparison with high radio luminosity objects we have used  similar
selection criteria to define a bright sample (3C*). This is described in 
the Appendix.

\begin{table*}
\caption{Complete listing of the samples}
\begin{tabular}{lllcccrrll}

Object & Samples & $z$&grade &$S_{38}$ & $S_{151}$& $R$ &Size&Class& Notes \\ 
       &         &    &      & /Jy     &  /Jy     &     &/kpc&     &       \\ 
1731+6641&7C-{\sc iii}              & 0.561&$\alpha$&0.9&0.52&21.4&7.4&HEG&\\  
1732+6535&7C-{\sc iii}, 8C-NEC, NEC*& 0.856& $\alpha$?&19.6&6.17&18.5&167&Q&\\
1732+6715& Temporarily excluded&   -   &   -      & 6.1&0.88&-   &-&-&Confused in radio \\
1733+6719&7C-{\sc iii}, 8C-NEC, NEC*& 1.84& $\beta$   &6.1&1.55&22.6&20.8&WQ&\\ 
1736+6504&8C-NEC, NEC*              & 2.40& $\alpha$  &2.2&0.48&23.0&133&NLG&\\
         &                      &     &           &   &    &    &   &\\
1736+6710&7C-{\sc iii}, 8C-NEC, NEC*&0.188& $\alpha$  &2.0&0.82&16.6&121&LEG&FRI\\
1740+6640&7C-{\sc iii}          & 2.10& $\alpha$  &0.9&0.54&24.8&$<$4.1&NLG&\\
1741+6704&7C-{\sc iii}, 8C-NEC, NEC*&1.054& $\alpha$  &1.9&0.72&22.8&33.4&NLG&\\
1742+6346&7C-{\sc iii}, 8C-NEC, NEC*&1.27&  $\beta$   &2.5&0.62&22.6&440&NLG&\\
1743+6431&7C-{\sc iii}, 8C-NEC, NEC*& ?  &   -        &8.2&1.89&23.3&-& ? &\\
         &                      &    &             &   &    &    &   &\\
1743+6344&7C-{\sc iii}, 8C-NEC, NEC*&0.324& $\alpha$  &5.5&1.59&21.3&80.5&LEG&ID uncertain\\
1743+6639&7C-{\sc iii}, 8C-NEC, NEC*&0.272& $\alpha$  &5.0&1.97&17.4&259&LEG&\\
1745+6415&7C-{\sc iii}, 8C-NEC, NEC*&0.673& $\alpha$  &2.1&0.59&20.5&44.2&HEG&\\
1745+6422&7C-{\sc iii}, 8C-NEC, NEC*&1.23& $\alpha$   &5.2&1.41&21.9&138&WQ &\\
1745+6624&7C-{\sc iii}          &3.01& $\beta$    &1.0&0.51& -  &$<$3.6&NLG&$K=20.2$\\
         &                      &    &             &   &    &    &    & \\
1747+6533&7C-{\sc iii}, 8C-NEC, NEC*&1.516& $\alpha$  &7.3&2.72&21.8&5.6&WQ &\\
1748+6703&7C-{\sc iii}, 8C-NEC, NEC*&  ?  &    -      &7.5&2.17&23.9&-& ? &\\
1748+6657&7C-{\sc iii}, NEC*    &1.045& $\beta$   &$<$0.8&1.15&22.4&2.6&NLG&\\
1748+6731&7C-{\sc iii}, 8C-NEC, NEC*&0.56& $\alpha$   &1.8&0.64&19.9&803&HEG?&\\
1751+6809&7C-{\sc iii}, 8C-NEC, NEC*&1.54& $\beta$    &2.4&1.03&22.9&17.1&NLG&\\
         &                      &    &             &   &    &    &   &\\
1750+6805&8C-NEC                    &  ? &    -       &1.3&0.35&$>$23.2&-& ? &8C1751+681C\\
1751+6741&8C-NEC, NEC*              &1.43& $\gamma$   &1.9&0.33&22.8&32.6&NLG&\\
1751+6455&7C-{\sc iii}, 8C-NEC, NEC*&0.294& $\alpha$   &1.8&0.65&17.1&234&LEG&\\
1753+6311&7C-{\sc iii}, 8C-NEC, NEC*&1.96&  $\gamma$   &4.4&1.06&22.6&140&NLG&\\ 
1753+6627&8C-NEC                    & ? &     -        &1.4&0.32&$>$23.6&-& ? &\\
         &                      &   &              &   &    &       &   &\\
1753+6543&7C-{\sc iii}, 8C-NEC, NEC*&0.140& $\alpha$   &4.6&1.62&17.2&272&WQ&\\
1754+6420&7C-{\sc iii}, 8C-NEC, NEC*&1.09&  $\beta$    &2.0&0.50&$>$22.4&129&NLG&\\
1754+6311&8C-NEC                    &1.23&  $\alpha$   &1.8&0.44&21.1&215&WQ &\\
1755+6314&7C-{\sc iii}, 8C-NEC, NEC*&0.388& $\alpha$   &4.8&1.19&18.0&254&LEG&FRI\\
1755+6830&7C-{\sc iii}, 8C-NEC, NEC*&0.744& $\alpha$   &3.8&1.11&20.0&72.1&NLG&\\
         &                      &     &            &   &    &    &   &\\
1756+6520&7C-{\sc iii}, 8C-NEC, NEC*&1.48&  $\gamma$   &2.6&0.67&$>$23.4&39.4&NLG&\\
1757+6741&8C-NEC                    &1.31&  $\beta$    &1.5&0.39&22.3&155&WQ&\\
1758+6535&7C-{\sc iii}, 8C-NEC, NEC*&0.80&  $\alpha$   &3.3&1.13&22.0&872&NLG\\
1758+6553&7C-{\sc iii}, 8C-NEC, NEC*&0.171& $\alpha$   &3.7&1.30&16.8&434&LEG&\\
1758+6307&7C-{\sc iii}, 8C-NEC, NEC*&1.19&  $\alpha$   &4.9&1.86&21.1&29.3&WQ&\\
         &                      &    &             &   &    &    &  &\\
1758+6719&7C-{\sc iii}, 8C-NEC, NEC*&2.70&  $\alpha$   &2.0&0.76&23.6&340&NLG&\\
1801+6902&7C-{\sc iii}, 8C-NEC, NEC*&1.27&  $\alpha$   &4.1&1.37&19.4&181&Q  &\\
1802+6456&7C-{\sc iii}, 8C-NEC, NEC*&2.11&  $\alpha$   &9.9&1.97&22.8&211&NLG&\\
1803+6605&8C-NEC, NEC*              &1.61&  $\alpha$   &2.3&0.44&21.1&306&WQ?&\\
1804+6625&7C-{\sc iii}, 8C-NEC      &1.91&  $\alpha$   &1.4&0.55&20.2&33.1&Q  &\\
                                &    &             &   &    &    &   &\\
1804+6313&7C-{\sc iii}, 8C-NEC, NEC*&  ? &    -        &3.1&0.62&23.0&-& ? &\\
1805+6332&7C-{\sc iii}, 8C-NEC, NEC*&1.84&  $\alpha$   &5.4&1.04&$>$23.7&117&NLG&\\
1807+6831&7C-{\sc iii}, 8C-NEC, NEC*&0.58&  $\alpha$   &7.0&2.12&19.9&218&HEG&\\
1807+6719&7C-{\sc iii}          &2.78&  $\alpha$   &1.0&0.71&$>$23.5&14.2&NLG&\\
1807+6841&7C-{\sc iii}, 8C-NEC, NEC*&0.816& $\alpha$   &2.3&0.60&22.0&99.2&NLG&\\
         &                      &     &            &   &    &    &   & \\
1807+6743&8C-NEC                    & ?   &   -        &2.0&0.36&21.6&-&?  &\\
1811+6321&7C-{\sc iii}, 8C-NEC, NEC*&0.273& $\gamma$   &3.0&0.95&19.3&270&LEG&\\
1812+6814&7C-{\sc iii}          &1.08 & $\alpha$   &1.2&0.59&22.5&189&NLG&\\
1813+6846&7C-{\sc iii}, 8C-NEC, NEC*&1.03 & $\alpha$   &4.6&1.51&19.3&444&Q &\\
1813+6439&7C-{\sc iii}, 8C-NEC, NEC*&2.04 & $\alpha$   &1.9&0.50&22.0&310&WQ&\\
         &                      &     &            &   &    &    &  &\\
1814+6702&7C-{\sc iii}, 8C-NEC, NEC*&4.05 & $\gamma$   &8.0&2.26&24.1&115&NLG&\\
1814+6550&8C-NEC                    &0.95 & $\beta$    &1.3&0.47&20.0&22.9&Q  &\\
1814+6529&7C-{\sc iii}, 8C-NEC, NEC*&0.96 & $\beta$    &4.7&1.22&22.8&1069&NLG&\\
1815+6805&7C-{\sc iii}, 8C-NEC, NEC*&0.230& $\alpha$   &5.9&1.96&17.8&233&WQ&\\
1815+6815&7C-{\sc iii}, 8C-NEC, NEC*&0.794& $\alpha$   &3.8&1.37&20.8&1643&NLG&\\
\end{tabular}
\end{table*}

\setcounter{table}{8}

\begin{table*}
\caption{Complete listing of the samples continued}
\begin{tabular}{lllcccrrll}

Object & Samples & $z$&grade &$S_{38}$ & $S_{151}$& $R$ &Size& Class& Notes \\ 
       &         &    &      & /Jy     &  /Jy     &     &/kpc&       \\     
1816+6710&7C-{\sc iii}, 8C-NEC, NEC*&0.92&  $\gamma$   &3.6&2.36&$>$23.8&16.0&NLG&$H=18.8$\\
1816+6605&7C-{\sc iii}, 8C-NEC, NEC*&0.92&  $\beta$    &3.8&1.29&22.3&228&NLG&\\
1819+6550&7C-{\sc iii}, 8C-NEC, NEC*&0.724& $\alpha$   &2.8&1.17&21.6&68.3&NLG&\\
1820+6657&7C-{\sc iii}          &2.98& $\gamma$    &$<$0.8&0.83&$>$24.1&$<$3.6&NLG&\\
1821+6442&Excluded& ?                &    -        &2.5&0.82&-&-&?&Bright star
at ID position\\
         &                      &    &             &   &    & & & \\
1821+6419A&8C-NEC, NEC*                &0.298&$\alpha$ &1.8&0.48&14.1&110&RQQ&E1821+643\\
1821+6419B&8C-NEC                      &0.304&$\alpha$? &1.6&0.40&18.1&36.7&LEG?&FRI in cluster with E1821+643\\
1822+6601&7C-{\sc iii}, 8C-NEC, NEC*&0.37& $\alpha$     &2.5&0.97&18.7&322&LEG&\\
1825+6602&7C-{\sc iii}, 8C-NEC, NEC*&2.38& $\gamma$     &2.1&0.84&23.4&$<$7.8&NLG&\\
1826+6510&7C-{\sc iii}, 8C-NEC, NEC*&0.646& $\alpha$    &3.0&1.39&20.7&265&LEG&\\
         &                      &      &            &   &    &    &   & \\ 
1826+6704&7C-{\sc iii}, 8C-NEC, NEC*&0.287& $\alpha$    &1.3&0.60&17.8&102&WQ?&\\
1827+6709&7C-{\sc iii}, 8C-NEC, NEC*&0.48& $\beta$      &2.7&1.10&21.0&119&LEG&\\
1827+6517&Excluded& ?      &        &3.1&0.78& -  &-&? &Bright star at ID position\\ 
\end{tabular}

Note: the $R$-magnitudes are measured in a 63-kpc metric aperture, limits are
3$\sigma$.
\end{table*}

\subsection{A complete listing of the samples}

In this subsection we present a complete listing of the 8C-NEC, 7C-{\sc iii}
and NEC* samples in a single table for ease of reference. Object 
classifications into low-excitation narrow-line radio galaxy (LEG), 
high-excitation narrow-line
radio galaxy (HEG), weak quasar (WQ) and quasar (Q) follow a combination of
Jackson \& Rawlings (1997) and Willott et al.\ (1998) criteria. For galaxies
with [O{\sc iii}]500.7 in the optical band ($z<0.7$ for most spectra) we 
have split the radio galaxies into LEG and HEG classes.
LEGs are defined to have [O{\sc iii}]500.7 EW $<1$ nm or 
[O{\sc ii}]372.7/[O{\sc iii}]500.7 $< 1$, 
all other narrow-line radio galaxies are HEGs. Where [O{\sc iii}] is redshifted
out of the optical band narrow-line radio galaxies 
are simply classified as NLG pending infrared spectroscopy. WQs have at least
one broad line observed, but $M_{\rm B}>-23$, whereas Qs have at least one 
broad line and $M_{\rm B}<-23$ in unresolved flux. To calculate $M_{\rm B}$ 
we have assumed 
a power-law quasar spectrum with an optical spectral index of 0.5 for
consistency with Willott et al.\ (1998).

\section{Correlations of FRII source properties}

A partial correlation analysis on the FRII or probable FRII sources 
from both the 7C-{\sc iii} + LRL and the NEC* + 3C* samples confirmed the same
main dependences as those found for the 7C-{\sc i}, {\sc ii}, 6C and 3C (LRL) 
samples studied by Blundell et al.\ (1999), namely that rest-frame spectral 
index at 1 GHz correlates best with radio luminosity, and that source size 
correlates best with redshift (also shown in Paper II). Our trends are 
consistent with those seen by Blundell et al., although as our 3C sample
is identical the results are not completely independent.

One difference between the 7C-{\sc iii} + LRL samples and the NEC* + 3C* 
samples is the lack of a strong spectral index -- size correlation 
in the latter (Fig.\ 8). This 
seems to be being driven mostly by the flattening of the spectra of the
CSS sources. 

\begin{figure*}
\begin{picture}(400,150)
\put(-30,180){\includegraphics{/data/arcturus/mdl/corr/nusnu/nec_adplot.ps}}
\put(220,180){\includegraphics{/data/arcturus/mdl/corr/7c3/7c3_adplot.ps}}
\put(-40,120){\Large \bf (a)}
\put(210,120){\Large \bf (b)}
\end{picture}
\caption{Size vs spectral index for (a) the NEC* + 3C* samples, and 
(b) the 7C-{\sc iii} + LRL samples. Crosses are 3C* in (a) and  LRL in 
(b), filled circles are NEC* in (a) and 7C-{\sc iii} in (b). Note the larger
number of small, flatter spectrum objects in the 7C-{\sc iii}+LRL sample 
compared to the NEC*+3C* sample.}
\end{figure*}

\subsection{Size evolution}

%\subsubsection{Background}

%\subsubsection{Choice of variables and correlation method}

We show in Fig.\ 9 the redshift-size relation for FRII sources from the 
7C-{\sc iii}+LRL and NEC*+3C* samples in two cosmologies. 
As is conventional we have parameterised the redshift dependence of source 
size in terms of the inverse scale factor $(1+z)$ as 
$D\propto (1+z)^{-\eta}$, and in Table 9 we give the best-fitting
exponents $\eta$ for all redshifts and, in the case of the NEC* + 3C* sample,
for the grade $\alpha$ redshifts only. We have also tried adding the three 
objects without spectroscopic redshifts in, with all redshifts set to either 
1.2 or 1.8, and find there is a small, but not significant 
reduction in $\eta$ of $<0.1$. The new selection 
criteria are effective at removing CSS objects, most 
of which seem to show a flattening in their spectra at low frequencies, and
therefore emit relatively low power at 151 MHz in the rest frame. This 
produces a ``cleaner'' looking correlation, indeed so good that 
we have felt confident using parametric methods to evaluate the slope
of the correlation. The slopes were determined using the 
Estimation-Maximisation method as implemented in the {\sc iraf} task 
{\sc emmethod} (Isobe \& Feigelson 1990)
which is able to deal with the size limits in the data by assuming a normal
distribution for the correlation residuals. As only one NEC* and four
7C-{\sc iii} objects had size limits this is unlikely to have affected the 
correlations. 

The correlation slopes for the NEC* + 3C* samples
are flatter than the Blundell et al.\ (1999) values. Though this is only 
marginally significant statistically, it is clear from examination of Fig.\ 9 
that this effect is due to larger numbers of high redshift CSS sources in 
the 151-MHz (observed-frame) selected samples. The small
size of these objects ($\stackrel{<}{_{\sim}} 30$kpc) means that most of them 
are probably confined by the interstellar medium of the host galaxy, and this 
high density environment will make the conversion of jet kinetic power into 
radio emission particularly efficient. A steep jet luminosity function 
(e.g.\ Kaiser \& Alexander 1998) will then lead to them being preferentially
included in flux-limited samples, particularly at high rest-frame
selection frequencies where their spectra are yet to start flattening.
For the 
7C-{\sc iii} sample we obtain slopes very similar to those of Blundell et al.\ 
(probably in part because we use the same LRL sample for
comparison). We thus find that the dependence of radio source size on redshift
is fairly weak, with radio source size evolving approximately with the scale 
factor of the Universe, $(1+z)^{-1}$. In particular the upper envelope of the 
size distribution is very weakly dependent on redshift.

The trend of decreasing size with redshift has usually been assumed
to be a straightforward consequence of increasing environmental density with 
redshift, combined with a finite source lifetime (e.g.\ Subramanian \& Swarup 
1990). However, the strong selection pressures associated with the steep radio 
luminosity function mean that what is being measured is the size 
at which, for a given redshift, the average radio source reaches its maximum
luminosity. Hence if, as argued by Blundell et al. (1999), high redshift, 
high luminosity 
sources reach their maximum luminosities earlier in their lives than low 
redshift ones due to a combination higher inverse-Compton losses and 
a correlation of injection spectral index with radio luminosity, this is 
also a possible 
explanation. Recently, Kaiser \& Alexander (1998) have also suggested that 
if either the lifetimes of radio sources or the shape of the density 
distribution of their environments affects the powers of the jets they 
can explain both the epoch dependence of linear size and the rapid evolution
in the luminosity function of the FRII population, at least within 3C. 
One observational way in which modelling uncertainties could be addressed
is by finding ``dead'' radio sources and estimating their ages; this may be 
the only way of constraining whether finite source lifetimes are likely to 
play an important r\^{o}le in evolutionary scenarios.

The influence of the
CSS sources in driving the redshift-size correlation in flux-density limited
samples of FRII radio sources 
could explain why much steeper slopes are seen when samples are 
selected at higher frequency, because these samples typically 
contain substantially larger fractions of CSS sources than LRL. The 
CSS sources could also account for the radio luminosity 
dependence too. The most likely mechanisms for spectral flattening 
(thermal absorption; Bicknell et al.\ 1997, or synchrotron self-absorption) 
are both likely to be luminosity dependent, thermal absorption through the 
emission line luminosity -- radio luminosity correlation and 
synchrotron self-absorption more directly. Kapahi (1989)  
removed CSS sources from his study of size evolution, and found that the 
best-fit value of $\eta$ was reduced (from $3.5 \pm 0.5$ to $2.5 \pm 0.5$) and
the dependence on luminosity was weakened ($\epsilon$ was reduced from 
$0.4 \pm 0.1$ to $0.2 \pm 0.1$). These correlations are 
still steeper than we observe for the NEC* + 3C* sample ($\eta=1.6 \pm 0.3$, 
$\epsilon \approx 0$), although the errors on the slopes both in our study and
that of Kapahi's are sufficient for the discrepancies to be due to a 
combination of statistical fluctuations and the lack of spectroscopic 
redshift completeness in the samples used by Kapahi in particular.

\begin{figure*}
\begin{picture}(500,150)
%\put(0,330){\special{psfile=/data/arcturus/mdl/corr/nusnu/nec_omeg1_dz.ps
%hscale=30 vscale=30 angle=-90}}
%\put(250,330){\special{psfile=/data/arcturus/mdl/corr/7c3/7c3_omeg1_dz.ps
%hscale=30 vscale=30 angle=-90}}
\put(0,170){\includegraphics{/data/arcturus/mdl/corr/nusnu/nec_lamp9_dz.ps}}
\put(250,170){\includegraphics{/data/arcturus/mdl/corr/7c3/7c3_lamp9_dz.ps}}
\put(0,120){\Large \bf (a)}
\put(250,120){\Large \bf (b)}
\end{picture}
\caption{The size-redshift relation (a) for sources in the NEC* and 3C* 
samples and (b) the 7C-{\sc iii} and LRL samples, plotted for 
$\Omega_{\rm M}=0.1$, $\Omega_{\rm \Lambda}=0.9$. Objects in the brighter 
samples are plotted as stars and objects in the fainter as squares. The dashed
lines show the best fitting lines from Table 6.}
\end{figure*}

\begin{table}
\caption{The lg$D$--lg$(1+z)$ correlations for the NEC* and 
7C-{\sc iii} samples}
\begin{tabular}{lll}
$\alpha$,$\beta$ \& $\gamma$ redshifts & & \\
Cosmology&\multicolumn{2}{l}{Correlation exponent, $\eta$}\\
         & NEC* + 3C*  & 7C-{\sc iii} + LRL \\
$\Omega_{\rm M} = 1$, $\Omega_{\Lambda}=0$&$1.60 \pm 0.29$&$2.09\pm 0.29$\\
$\Omega_{\rm M} = 0$, $\Omega_{\Lambda}=0$&$1.17 \pm 0.29$&$1.66 \pm 0.29$\\
$\Omega_{\rm M} = 0.1$, $\Omega_{\Lambda}=0.9$&$1.07 \pm 0.29$&$1.57 
\pm 0.29$\\
                        &&\\
$\alpha$ redshifts only && \\
Cosmology&\multicolumn{2}{l}{Correlation exponent, $\eta$}\\
         & NEC* + 3C*  &  \\
$\Omega_{\rm M} = 1$, $\Omega_{\Lambda}=0$&$1.29 \pm 0.30$&\\
$\Omega_{\rm M} = 0$, $\Omega_{\Lambda}=0$&$0.88 \pm 0.30$&\\
$\Omega_{\rm M} = 0.1$, $\Omega_{\Lambda}=0.9$&$0.74 \pm 0.30$&\\ 
\end{tabular}

\vspace*{0.1in}

%\noindent
\end{table}

\section*{acknowledgements}
We would like to thank Christian Kaiser and Jasper Wall for their helpful 
comments  on the manuscript, and Malcolm Bremer, Rob van Ojik 
and Richard Saunders for assistance during the 1993 June WHT observations.
We also thank the staff at the WHT, McDonald Observatory, Lick Observatory and 
the IRTF for their assistance. 
The WHT is operated on the island of La Palma by
PPARC in the Spanish Observatorio del
Roque de los Muchachos of the Instituto de Astrofisica de
Canarias. SER was a visiting astronomer at the NASA Infrared Telescope 
Facility, operated by the University of Hawaii under contract with NASA.
Offset star positions were obtained using astrometry from the guide stars 
selection system astrometric support system (GASP), developed at the Space Telescope Science Institute, which is operated by the Association of Universities for Research in Astronomy, Inc.\, for NASA.
This research has made use of the NASA/IPAC Extragalactic Database 
(NED) which is operated by the Jet Propulsion Laboratory, California 
Institute of Technology, under contract with the National Aeronautics and Space
Administration. 

\appendix

%\section[]{Additional imaging data on the 7C-{\sc iii} sample}

\section[]{The 3C* sample}

To select a bright comparison sample in a similar manner to the NEC* sample 
we used data from two 38 MHz surveys to obtain low frequency flux
densities for radio sources in the sample of LRL, resulting in two 
different flux density limits. In the region covered 
by the 38-MHz survey of Rees (1990) we used $S_{151\; {\rm (rest)}}/(1+z) 
\geq 12$ Jy, and in the remainder of the northern sky, 
$S_{151\; {\rm (rest)}}/(1+z) \geq 15$ Jy
using flux density from the 38-MHz survey of Williams, Kenderdine \& Baldwin 
(1966) corrected according to Laing \& Peacock (1980). Flux densities at 
higher frequencies were mostly obtained from Laing \& Peacock (1980).
The members of the sample are listed in Table A1, and 
those of the NEC* sample in Table A2.

The redshift distributions of the LRL and 3C* samples are not significantly 
different statistically; a Kolmogorov-Smirnov test on the two samples gives
a probability of 18 per cent that they are drawn from the same underlying 
distribution. There does, however, seem to be a relative lack of high 
redshift objects 
in the 3C* sample due to the CSS sources dropping out (Fig.\ A1). 

\begin{figure}
\begin{picture}(200,220)
\put(-10,-80){\includegraphics{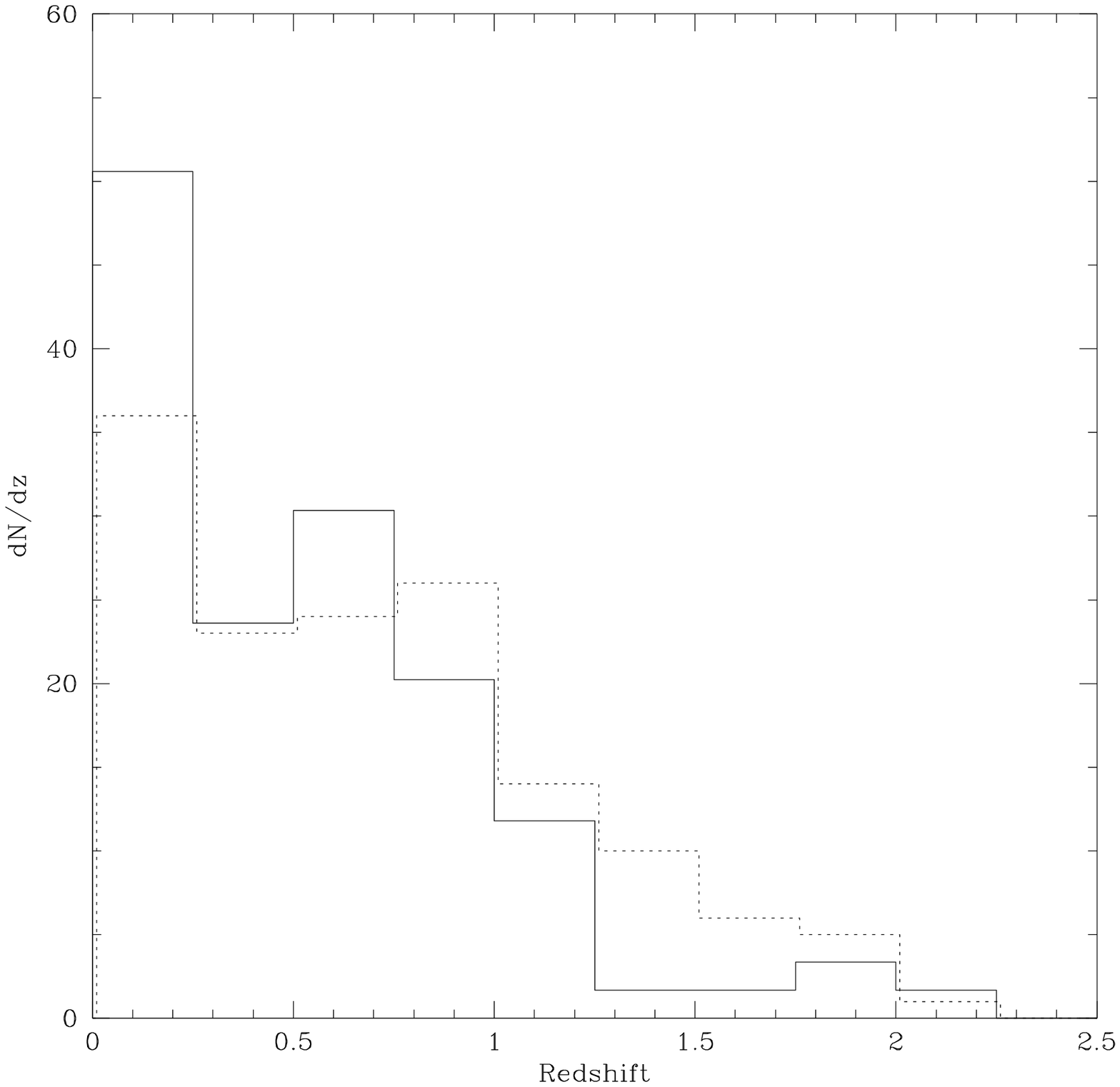}}
\end{picture}
\caption{Redshift distributions for LRL (dashed) and 3C* (solid). Note that the
3C* redshift distribution has been multiplied by 1.7 to normalise to the 
number of objects in LRL.}
\end{figure}

\begin{table}
\caption{The FRII sources in the 3C* sample with rest-frame flux densities
at 151 MHz and rest-frame spectral indices at 1 GHz}
\begin{tabular}{lrrrcr}
Name  & redshift &$S_{151\; {\rm (rf)}}$&$\alpha_{1 {\rm GHz\; (rf)}}$&Type&radio size \\
      &          &  /Jy     &    &  & /arcsec\\
3C6.1  &  .8404  &   25.6&0.77&HEG&   26.0\\	    
3C9    & 2.0120  &   63.8&1.04&Q&   14.0\\	    
3C20   &  .1740  &   59.5&0.75&HEG&   53.0\\	    
3C28   &  .1952  &   27.1&1.18&LEG&   52.0\\	    
3C33   &  .0595  &   66.8&0.73&HEG&  253.0\\	    
3C33.1 &  .1810  &   19.9&0.86&WQ&  227.0\\	    
3C34   &  .6900  &   28.4&1.04&HEG&   48.0\\	    
3C35   &  .0670  &   17.5&0.88&LEG&  730.0\\	    
3C47   &  .4250  &   50.1&0.99&Q&   69.0\\	    
3C55   &  .7350  &   42.7&0.98&HEG?&   71.0\\	    
3C61.1 &  .1860  &   42.8&0.83&LEG&  185.0\\	    
3C68.2 & 1.5750  &   47.9&1.23&HEG?&   22.3\\	    
3C79   &  .2559  &   42.7&0.95&HEG&   87.0\\	    
3C98   &  .0306  &   56.6&0.69&HEG&  310.0\\	    
3C109  &  .3056  &   32.0&0.87&WQ&   90.0\\	    
3C123  &  .2177  &  290.3&0.78&LEG&   40.0\\	    
3C132  &  .2140  &   21.0&0.80&LEG&   23.0\\	    
3C147  &  .5450  &   74.6&0.57&Q&    3.0\\	    
3C153  &  .2769  &   22.9&0.76&HEG&    8.5\\	    
3C172  &  .5191  &   24.8&0.86&HEG&  103.0\\	    
3C173.1&  .2920  &   23.5&0.91&LEG&   60.0\\	    
3C175  &  .7680  &   41.2&1.03&Q&   48.0\\	    
3C184  &  .9900  &   24.9&0.87&HEG&    4.8\\	    
3C184.1&  .1182  &   18.1&0.76&HEG&  167.0\\	    
3C186  & 1.0630  &   34.9&1.14&Q&  100.0\\	    
DA240  &  .0350  &   29.8&0.83&LEG& 2164.0\\	    
3C192  &  .0598  &   29.6&0.74&HEG&  196.0\\	    
3C196  &  .8710  &  124.3&0.80&Q&   10.0\\	    
3C200  &  .4580  &   22.1&0.94&LEG&   26.0\\	    
3C204  & 1.1120  &   28.8&1.08&Q&   35.0\\	    
3C207  &  .6840  &   26.4&0.83&Q&   14.0\\	    
3C208  & 1.1100  &   43.2&1.05&Q&   11.0\\	    
3C216  &  .6680  &   38.2&0.99&Q&   30.0\\	    
3C219  &  .1744  &   59.3&0.89&WQ&  189.0\\	    
3C220.1&  .6100  &   28.8&1.01&HEG&   30.0\\	    
3C220.3&  .6850  &   30.6&0.95&HEG&    7.4\\	    
3C223  &  .1368  &   20.4&0.76&HEG&  303.0\\	    
3C225B &  .5820  &   32.0&0.92&HEG?&    4.6\\	    
3C226  &  .8177  &   31.1&0.95&HEG&   35.0\\	    
4C73.08&  .0581  &   19.8&0.92&HEG&  947.0\\	    
3C228  &  .5520  &   32.3&0.85&HEG&   45.0\\	    
3C234  &  .1848  &   48.8&0.96&WQ&  110.0\\	    
3C239  & 1.7900  &   44.5&1.07&HEG&   11.2\\	    
3C244.1&  .4280  &   36.1&0.91&HEG&   53.0\\	    
3C249.1&  .3110  &   18.6&0.87&Q&   44.0\\	    
3C252  & 1.1050  &   35.7&1.12&HEG&   60.0\\	    
3C254  &  .7340  &   38.2&0.96&Q&   13.1\\	    
3C263  &  .6563  &   28.1&0.85&Q&   44.2\\	    
3C263.1&  .8240  &   33.6&0.90&HEG&    6.8\\	    
3C265  &  .8108  &   44.6&1.02&HEG&   78.0\\	    
3C267  & 1.1420  &   32.6&0.90&HEG&   38.0\\	    
3C268.1&  .9737  &   39.3&0.64&HEG&   46.0\\	    
3C274.1&  .4220  &   28.7&0.94&HEG&  150.0\\	    
3C275.1&  .5570  &   30.5&0.94&Q&   18.0\\	    
3C277.2&  .7660  &   26.7&0.96&HEG&   58.0\\	    
3C280  &  .9960  &   46.0&0.79&HEG&   14.5\\	    
3C292  &  .7130  &   22.7&0.86&HEG&  133.0\\	    
3C293  &  .0452  &   20.5&0.69&LEG?&  216.0\\	    
3C294  & 1.7800  &   45.2&1.11&HEG&   15.0\\	    
3C295  &  .4614  &   98.8&0.70&HEG&    5.3\\	    
3C300  &  .2700  &   29.4&0.88&HEG&  100.0\\	    
3C309.1&  .9040  &   39.4&0.76&Q&    2.9\\	    
3C310  &  .0540  &   86.3&1.20&LEG?&  310.0\\	    
3C319  &  .1920  &   22.1&0.97&LEG&  105.0\\	    
\end{tabular}

\end{table}

\setcounter{table}{0}

\begin{table}
\caption{continued}
\begin{tabular}{lrrrcr}
Name  & redshift &$S_{151\; {\rm (rf)}}$&$\alpha_{1 {\rm GHz\; (rf)}}$&Type&radio size \\
       &         &  /Jy     &    &  & /arcsec   \\
3C321  &  .0960  &   23.7&0.84&HEG &  290.0\\	    
3C324  & 1.2063  &   37.3&0.93&HEG &   10.0\\	    
3C326  &  .0895  &   26.3&0.84&LEG & 1190.0\\	    
3C325  &  .8600  &   27.6&0.80&Q &   16.0\\	    
3C330  &  .5490  &   48.4&0.77&HEG &   62.0\\	    
3C334  &  .5550  &   25.3&0.95&Q &   58.0\\	    
3C349  &  .2050  &   20.7&0.79&HEG &   82.0\\	    
3C351  &  .3710  &   21.5&0.75&Q &   65.0\\	    
4C13.66& 1.4500  &   40.2&1.07&HEG &    6.0\\	    
3C368  & 1.1320  &   41.6&1.24&HEG &    7.9\\	    
3C380  &  .6910  &  127.1&0.83&Q &   20.0\\	    
3C381  &  .1605  &   21.0&0.76&HEG &   69.0\\	    
3C382  &  .0578  &   27.6&0.74&WQ &  189.0\\	    
3C388  &  .0908  &   33.5&0.84&HEG &   43.0\\	    
3C390.3&  .0569  &   62.4&0.77&WQ &  215.0\\	    
3C401  &  .2010  &   29.3&0.84&LEG &   24.0\\	    
3C427.1&  .5720  &   49.4&0.99&LEG &   23.1\\	    
3C436  &  .2145  &   26.3&0.89&HEG &  104.0\\	    
3C438  &  .2900  &   68.9&1.02&HEG &   22.0\\	    
3C442A &  .0263  &   26.3&0.99&LEG &  590.0\\	    
3C452  &  .0811  &   73.5&0.89&HEG &  250.0\\	    
3C457  &  .4270  &   23.4&1.03&HEG &  190.0\\
\end{tabular}
\end{table}

\begin{table}
\caption{The FRII and probable FRII sources in the NEC* sample with 
rest-frame 151 MHz flux densities and rest-frame 1 GHz spectral indices}
\begin{tabular}{lrrrr}
Name  & redshift &$S_{151\; {\rm (rf)}}$&$\alpha_{1 {\rm GHz\; (rf)}}$&radio size \\
          &     & /Jy &      & /arcsec\\
 1732+6535&0.856& 9.87&0.82 &  20.0\\ 
 1733+6719&1.84 & 4.46&0.87 &   2.5\\ 
 1736+6504&2.40 & 2.25&1.28 &  17.0\\ 
%1736+6710&0.188& 0.95&0.94 &  30.0\\
 1741+6704&1.054& 1.23&0.96 &   3.9\\ 
 1742+6346&1.27 & 1.41&0.97 &  51.0\\
 1743+6344&0.324& 1.79&0.83 &  14.0\\
 1743+6431&  -  &  -  &  -  &  45.0\\
 1743+6639&0.272& 2.39&0.71 &  50.0\\
 1745+6415&0.673& 1.00&0.69 &   5.6\\
 1745+6422&1.23 & 3.27&0.95 &  16.0\\ 
 1747+6533&1.516& 5.80&0.85 &   0.7\\
 1748+6703&  -  &  -  & -   &  14.0\\
 1748+6657&1.045& 1.24&0.42 &   0.3\\ 
 1748+6731&0.56 & 0.91&0.78 & 108.0\\ 
 1751+6809&1.54&  1.68&0.77 &   2.0\\ 
 1751+6741&1.43 & 1.13&1.39 &   3.8\\ 
 1751+6455&0.294& 0.86&0.79 &  43.0\\
 1753+6311&1.96 & 3.16&0.96 &  17.0\\
 1753+6543&0.140& 1.93&0.67 &  84.0\\
 1754+6420&1.09 & 0.99&0.88 &  15.0\\
%1755+6314&0.388& 1.68&0.82 &  40.0\\ 
 1755+6830&0.744& 1.97&0.87 &   8.9\\ 
 1756+6520&1.48 & 1.48&0.79 &   4.6\\
 1758+6535&0.80 & 1.81&0.74 & 106.0\\
 1758+6553&0.171& 1.54&0.90 & 115.0\\
 1758+6307&1.19 & 3.39&0.81 &   3.4\\ 
 1758+6719&2.70 & 1.91&0.85 &  45.0\\ 
 1801+6902&1.27 & 2.75&0.97 &  21.0\\ 
 1802+6456&2.11 & 6.67&1.03 &  26.0\\ 
 1803+6605&1.61 & 1.30&1.07 &  36.0\\
 1804+6313& -   &  -  &  -  &  29.0\\ 
 1805+6332&1.84 & 3.50&0.96 &  14.0\\ 
 1807+6831&0.58 & 3.27&0.84 &  29.0\\   
 1807+6841&0.816& 1.09&0.82 &  12.0\\   
 1811+6321&0.273& 1.12&0.87 &  52.0\\   
 1813+6846&1.03 & 2.67&0.88 &  52.0\\   
 1813+6439&2.04 & 1.58&1.07 &  38.0\\   
 1814+6702&4.05 &10.98&0.98 &  18.0\\   
 1814+6529&0.96 & 2.44&1.02 & 126.0\\   
 1815+6805&0.230& 2.50&0.91 &  50.0\\   
 1815+6815&0.794& 2.12&0.93 & 200.0\\   
 1816+6710&0.92 & 3.00&0.75 &  27.0\\   
 1816+6605&0.92 & 2.24&0.93 &   1.9\\    
 1819+6550&0.724& 1.71&0.73 &   8.5\\    
%1821+641A&0.298& 0.61&1.22 &  21.0\\    
 1822+6601&0.37 & 1.29&0.81 &  52.0\\   
 1825+6602&2.38 & 2.72&0.83 &$<$1.0\\   
 1826+6510&0.63 & 1.91&0.83 &  34.0\\   
 1826+6704&0.287& 0.71&0.75 &  19.0\\   
 1827+6709&0.48 & 1.45&0.85 &  17.0\\   
\end{tabular}
\end{table}

 \end{document}